\documentclass[11pt,preprint]{aastex}
\usepackage{color}
\usepackage{courier}
\usepackage{lscape}
\usepackage{subfigure}
\usepackage{graphicx}

\def\simlt{\lower.5ex\hbox{$\; \buildrel < \over \sim \;$}}
\def\simgt{\lower.5ex\hbox{$\; \buildrel > \over \sim \;$}}

\def\gsim{\lower 2pt \hbox{$\, \buildrel {\scriptstyle >}\over
{\scriptstyle \sim}\,$}}
\def\lsim{\lower 2pt \hbox{$\, \buildrel {\scriptstyle <}\over
{\scriptstyle \sim}\,$}}

\def\deg{\ifmmode ^{\circ}
         \else $^{\circ}$\fi}
\def\pdeg{\ifmmode
           $\setbox0=\hbox{$^{\circ}$}\rlap{\hskip.11\wd0 .}$^{\circ}
     \else \setbox0=\hbox{$^{\circ}$}\rlap{\hskip.11\wd0 .}$^{\circ}$\fi}
     
\def\pc{\ifmmode \mathrm{pc} \else $\mathrm{pc}$ \fi}
\def\mpc{\ifmmode \mathrm{Mpc} \else $\mathrm{Mpc}$\fi}
\def\mpcthree{\ifmmode \mathrm{Mpc}^{-3} \else $\mathrm{Mpc}^{-3}$\fi}
\def\gpcthree{\ifmmode \mathrm{Gpc}^{-3} \else $\mathrm{Gpc}^{-3}$\fi}

\def\kelvin{\ifmmode \mathrm{K} \else {$\mathrm{K}$}\fi}
\def\kev{\ifmmode \mathrm{keV} \else $\mathrm{keV}$ \fi}

\def\lsun{\ifmmode {L_\odot} \else $L_\odot$\fi}
\def\msun{\ifmmode M_\odot \else $M_\odot$\fi}
\def\msunyr{\ifmmode M_\odot~\mathrm{yr}^{-1} \else $M_\odot~\mathrm{yr}^{-1}$\fi}

\def\cosi{\ifmmode {\cos\,i} \else $\cos\,i$\fi}

\def\heii{\ifmmode {\rm He{\sc ii}} \else He~{\sc ii}\fi}
\def\mgii{\ifmmode {\rm Mg{\sc ii}} \else Mg~{\sc ii}\fi}
\def\caii{\ifmmode {\rm Ca{\sc ii}} \else Ca~{\sc ii}\fi}
\def\ciii{\ifmmode {\rm C{\sc iii}]} \else C~{\sc iii}]\fi}
\def\civ{\ifmmode {\rm C{\sc iv}} \else C~{\sc iv}\fi}
\def\mgii{\ifmmode {\rm Mg{\sc ii}} \else Mg~{\sc ii}\fi}
\newcommand{\oi}{{\sc [O~i]}}

\newcommand{\oiii}{{\sc [O~iii]}}
\newcommand{\oiv}{{\sc [O~iv]}}

\newcommand{\nii}{{\sc [N~ii]}}

\newcommand{\neii}{{[Ne~{\sc ii}]}}
\newcommand{\nev}{{[Ne~{\sc v}]}}

\newcommand{\sii}{{\sc [S~ii]}}
\newcommand{\feii}{[Fe~{\sc ii}]}

\def\teff{\ifmmode {T_{\rm eff}} \else $T_{\rm eff}$\fi}
\def\tmax{\ifmmode {T_{\rm max}} \else $T_{\rm max}$\fi}

\def\mbh{\ifmmode {M_{\rm BH}} \else $M_{\rm BH}$\fi}
\def\led{\ifmmode L_{\mathrm{Ed}} \else $L_{\mathrm{Ed}}$\fi}
\def\lbolflare{\ifmmode L_{\mathrm{bol,flare}} \else $L_{\mathrm{bol,flare}}$\fi}
\def\lagn{\ifmmode L_{\mathrm{agn}} \else $L_{\mathrm{agn}}$\fi}
\def\lbolagn{\ifmmode L_{\mathrm{bol,agn}} \else $L_{\mathrm{bol,agn}}$\fi}
\def\lbol{\ifmmode L_{\mathrm{bol}} \else $L_{\mathrm{bol}}$\fi}
\def\mdot{\ifmmode {\dot M} \else $\dot M$\fi}
\def\mdoto{\ifmmode {\dot{M}_0} \else  $\dot{M}_0$\fi}
\def\mdotf{\ifmmode {\dot{M}_\mathrm{flare}} \else  $\dot{M}_\mathrm{flare}$\fi}

\def\hnot{\ifmmode H_0 \else H$_0$ \fi}

\def\vkep{\ifmmode v_\mathrm{Kep} \else $v_\mathrm{Kep}$ \fi}
\def\vc{\ifmmode v_\mathrm{c} \else $v_\mathrm{c}$ \fi}

\def\vthree{\ifmmode v_{1000} \else $v_{1000}$ \fi}
\def\vrel{\ifmmode v_\mathrm{rel} \else $v_\mathrm{rel}$ \fi}
\def\vkick{\ifmmode v_\mathrm{kick} \else $v_\mathrm{kick}$ \fi}
\def\vkickz{\ifmmode v_{\mathrm{kick},z} \else $v_{\mathrm{kick},z} $ \fi}
\def\vkicky{\ifmmode v_{\mathrm{kick},y} \else $v_{\mathrm{kick},y} $ \fi}
\def\vchar{\ifmmode v_\mathrm{char} \else $v_\mathrm{char}$ \fi}
\def\eflare{\ifmmode E_\mathrm{flare} \else $E_\mathrm{flare}$ \fi}
\def\ekick{\ifmmode E_\mathrm{kick} \else $E_\mathrm{kick}$ \fi}
\def\ecoll{\ifmmode E_\mathrm{coll} \else $E_\mathrm{coll}$ \fi}
\def\ezero{\ifmmode E_\mathrm{0} \else $E_\mathrm{0}$ \fi}
\def\efac{\ifmmode \xi_\mathrm{E} \else $\xi_\mathrm{E}$ \fi}
\def\tqso{\ifmmode t_\mathrm{QSO} \else $t_\mathrm{QSO}$ \fi}
\def\tflare{\ifmmode t_\mathrm{flare} \else $t_\mathrm{flare}$ \fi}
\def\tzero{\ifmmode t_\mathrm{0} \else $t_\mathrm{0}$ \fi}
\def\tfac{\ifmmode \xi_\mathrm{t} \else $\xi_\mathrm{t}$ \fi}
\def\gfac{\ifmmode f_\mathrm{g} \else $f_\mathrm{g}$ \fi}
\def\lflare{\ifmmode L_\mathrm{flare} \else $L_\mathrm{flare}$ \fi}
\def\fflare{\ifmmode F_\mathrm{flare} \else $F_\mathrm{flare}$ \fi}
\def\nflare{\ifmmode N_\mathrm{flare} \else $N_\mathrm{flare}$ \fi}
\def\tshock{\ifmmode T_\mathrm{shock} \else $T_\mathrm{shock}$ \fi}
\def\rmin{\ifmmode R_\mathrm{1} \else $R_\mathrm{1}$ \fi}
\def\rmax{\ifmmode R_\mathrm{2} \else $R_\mathrm{2}$ \fi}
\def\rbound{\ifmmode R_\mathrm{b} \else $R_\mathrm{b}$ \fi}
\def\pbound{\ifmmode P_\mathrm{b} \else $P_\mathrm{b}$ \fi}
\def\mbound{\ifmmode M_\mathrm{b} \else $M_\mathrm{b}$ \fi}
\def\mbo{\ifmmode M_{\mathrm{b}0} \else $M_{\mathrm{b}0} $ \fi}
\def\ebo{\ifmmode E_{\mathrm{b}0} \else $E_{\mathrm{b}0} $ \fi}
\def\efinal{\ifmmode E_\mathrm{final} \else $E_\mathrm{final} $ \fi}
\def\tbound{\ifmmode t_\mathrm{b} \else $t_\mathrm{b}$ \fi}
\def\tagn{\ifmmode t_\mathrm{AGN} \else $t_\mathrm{AGN}$ \fi}
\def\torb{\ifmmode t_\mathrm{orb} \else $t_\mathrm{orb}$ \fi}
\def\tdf{\ifmmode t_\mathrm{df} \else $t_\mathrm{df}$ \fi}
\def\rlim{\ifmmode R_\mathrm{lim} \else $R_\mathrm{lim}$ \fi}
\def\vlim{\ifmmode v_\mathrm{lim} \else $v_\mathrm{lim}$ \fi}
\def\vphi{\ifmmode v_\phi \else $v_\phi$ \fi}
\def\mlim{\ifmmode M_\mathrm{lim} \else $M_\mathrm{lim}$ \fi}
\def\tlim{\ifmmode t_\mathrm{lim} \else $t_\mathrm{lim}$ \fi}
\def\llim{\ifmmode L_\mathrm{lim} \else $L_\mathrm{lim}$ \fi}
\def\fqso{\ifmmode f_\mathrm{QSO} \else $f_\mathrm{QSO}$ \fi}

\def\hbeta{\ifmmode \rm{H}\beta \else H$\beta$\fi}
\def\hbetan{\ifmmode \rm{H}\beta_{\rm n} \else H$\beta_{\rm n}$\fi}
\def\hgamma{\ifmmode \rm{H}\gamma \else H$\gamma$\fi}
\def\hdelta{\ifmmode \rm{H}\delta \else H$\delta$\fi}
\def\hepsilon{\ifmmode \rm{H}\epsilon \else H$\epsilon$\fi}
\def\hzeta{\ifmmode \rm{H}\zeta \else H$\zeta$\fi}
\def\halpha{\ifmmode \rm{H}\alpha \else H$\alpha$\fi}
\def\lalpha{\ifmmode \rm{Ly}\alpha \else Ly$\alpha$}

\def\dvhb{\ifmmode \Delta v_{\hbeta} \else $\Delta v_{\hbeta}$\fi}
\def\dvmg{\ifmmode \Delta v_{\rm{Mg}} \else $\Delta v_{\rm{Mg}}$\fi}

\def\muobs{\ifmmode {\mu_{o}} \else  $\mu_{o}$ \fi}
\def\cosi{\ifmmode {\mathrm{cos}\,i} \else $\mathrm{cos}\,i$\fi}

\def\teff{\ifmmode {T_{eff}} \else $T_{eff}$ \fi}
\def\tmax{\ifmmode {T_{max}} \else $T_{max}$ \fi}

\def\tauh{\ifmmode {\tau_{\rm H}} \else $\tau_{\rm H}$ \fi}

\def\yr{\ifmmode {\rm yr} \else  yr \fi}
\def\kms{\ifmmode \rm km~s^{-1}\else $\rm km~s^{-1}$\fi}
\def\cm{\ifmmode {\rm cm} \else  cm \fi}
\def\cmmitwo{\ifmmode \rm cm^{-2} \else $\rm cm^{-2}$\fi}
\def\cmmithree{\ifmmode \rm cm^{-3} \else $\rm cm^{-3}$\fi}
\def\cmps{\ifmmode \rm cm~s^{-1}\else $\rm cm~s^{-1}$\fi}
\def\cmpsps{\ifmmode \rm cm~s^{-2}\else $\rm cm~s^{-2}$\fi}
\def\kmps{\ifmmode \rm km~s^{-1}\else $\rm km~s^{-1}$\fi}
\def\kmpspmpc{\ifmmode \rm km~s^{-1}~Mpc^{-1} \else
    $\rm km~s^{-1}~Mpc^{-1}$\fi}
  
\def\gcmthree{\ifmmode \rm g~cm^{-3} \else $\rm g~cm^{-3}$\fi}
\def\gcmtwo{\ifmmode \rm g~cm^{-2} \else $\rm g~cm^{-2}$\fi}
   
\def\erg{\ifmmode {\rm erg} \else $\rm erg$ \fi}
\def\ergps{\ifmmode {\rm erg~s^{-1}} \else $\rm erg~s^{-1}$ \fi}
\def\ergcms{\ifmmode \rm erg~cm^{-2}~s^{-1} \else $\rm erg~cm^{-2}~s^{-1}$ \fi}
\def\ergcmshz{\ifmmode \rm erg~s^{-1}~cm^{-2}~Hz^{-1} \else $\rm
erg~cm^{-2}~s^{-1}~Hz^{-1}$ \fi}
\def\ergcmsa{\ifmmode \rm erg~cm^{-2}~s^{-1}~\AA^{-1} \else $\rm
erg~cm^{-2}~s^{-1}~\AA^{-1}$ \fi}
\def\ergshz{\ifmmode \rm erg s^{-1} Hz^{-1} \else
   $\rm erg s^{-1} Hz^{-1}$ \fi}

\def\lam{\ifmmode {\lambda} \else {$\lambda$} \fi}
\def\llam{\ifmmode {L_\lambda} \else  $L_\lambda$ \fi}
\def\lamLlam{\ifmmode \lambda L_{\lambda}(5100) \else {$\lambda L_{\lambda}(5100)$} \fi}
\def\nuLnu{\ifmmode \nu L_{\nu}(5100) \else {$\nu L_{\nu}(5100)$} \fi}
\def\ilam{\ifmmode {I_\lambda} \else  $I_\lambda$ \fi}
\def\flam{\ifmmode {F_\lambda} \else  $F_\lambda$ \fi}
\def\inu{\ifmmode {I_\nu} \else  $I_\nu$ \fi}
\def\fnu{\ifmmode {F_\nu} \else  $F_\nu$ \fi}
\def\bnu{\ifmmode {B_\nu} \else  $B_\nu$ \fi}

\def\msigma{\ifmmode M_{\sigma} \else $M_{\sigma}$\fi}
\def\mbulge{\ifmmode M_{\mathrm{bulge}} \else $M_{\mathrm{bulge}}$\fi}
\def\mgal{\ifmmode M_{\mathrm{gal}} \else $M_{\mathrm{gal}}$\fi}
\def\lgal{\ifmmode L_{\mathrm{gal}} \else $L_{\mathrm{gal}}$\fi}
\def\lbulge{\ifmmode L_{\mathrm{bulge}} \else $L_{\mathrm{bulge}}$\fi}
\def\mgalstar{\ifmmode M^*_{\mathrm{gal}} \else $M^*_{\mathrm{gal}}$\fi}

\def\mbhsigstar{\ifmmode M_{\mathrm{BH}} - \sigma_* \else $M_{\mathrm{BH}} - \sigma_*$ \fi}
\def\deltalogmbh{\ifmmode \Delta~{\mathrm{log}}~M_{\mathrm{BH}} \else $\Delta$~log~$M_{\mathrm{BH}}$\fi}

\def\sigstar{\ifmmode \sigma_* \else $\sigma_*$\fi}
\def\sigthree{\ifmmode \sigma_{\mathrm{[O~III]}} \else $\sigma_{\mathrm{[O~III]}}$\fi}
\def\sigtwo{\ifmmode \sigma_{\mathrm{[O~II]}} \else $\sigma_{\mathrm{[O~II]}}$\fi}
\def\signl{\ifmmode \sigma_{\mathrm{NL}} \else $\sigma_{\mathrm{NL}}$\fi}
\def\wthree{\ifmmode {\rm FWHM({[O~III]})} \else $FWHM({[O~III]})$ \fi}
\def\wtwo{\ifmmode {\rm FWHM({[O~II]})} \else $FWHM({[O~II]})$ \fi}
\def\mthree{\ifmmode M_{\mathrm [O~III]} \else $M_{\mathrm [O~III]}$ \fi}
\def\mtwo{\ifmmode M_{\mathrm [O II]} \else $M_{\mathrm [O II]}$ \fi}

\def\lbreak{\ifmmode L_{\mathrm{break}} \else $L_{\mathrm{break}}$\fi}
\def\lcut{\ifmmode L_{\mathrm{cut}} \else $L_{\mathrm{cut}}$\fi}

\shortauthors{Smith, Koss \& Mushotzky}
\shorttitle{XBONGs and Optically Elusive AGN}

\begin{document}

\title{An Infrared and Optical Analysis of a Sample of XBONGs and Optically-Elusive AGN}

\author{K. L. Smith\altaffilmark{1}, M.~Koss\altaffilmark{2} \& R.F.~Mushotzky\altaffilmark{1}}

\altaffiltext{1}{Department of Astronomy, University of Maryland College Park; klsmith@astro.umd.edu, richard@astro.umd.edu}

\altaffiltext{2}{Institute for Astronomy, Department of Physics, ETH Zurich, Wolfgang-Pauli-Strasse 27, CH-8093 Zurich, Switzerland, mike.koss@phys.ethz.ch}

\begin{abstract}

We present near-infrared (NIR) spectra of four optically-elusive AGN and four X-ray bright, optically normal galaxies (XBONGs) from the \emph{Swift}-BAT survey. With archival observations from SDSS, 2MASS, \emph{Spitzer} and WISE, we test a number of AGN indicators in the NIR and MIR; namely, NIR emission line diagnostic ratios, the presence of coronal high-ionization lines, and infrared photometry. Of our eight hard X-ray selected AGN, we find that optical normalcy has a variety of causes from object to object, and no one explanation applies. Our objects have normal Eddington ratios, and so are unlikely to host radiatively-inefficient accretion flows (RIAFs). It is unlikely that star formation in the host or starlight dilution is contributing to their failure of optical diagnostics, except perhaps in two cases. The NIR continua are well-fit by two blackbodies: one at the stellar temperature, and a hot dust component near the dust sublimation temperature. XBONGs are more likely to have significant hot dust components, while these components are small relative to starlight in the optically-elusive AGN. Some of our sample have NIR line ratios typical of AGN, but NIR diagnostics are unsuccessful in distinguishing HII regions from AGN in general. In one object, we discover a hidden broad line region in the NIR. These results have strong relevance to the origin of optically normal AGN in deep X-ray surveys.

\end{abstract}


\section{Introduction}
\label{sec:intro}

Since the discovery of the X-ray background (XRB) in 1963, numerous evolving efforts have been made to resolve its sources. The ROSAT satellite was able to resolve most of the soft X-ray background into unobscured quasars. However, the overall spectral energy distribution of the background as a whole was not well approximated until the deep imaging capability of \emph{Chandra} resolved the hard XRB into point sources: quasars and active galactic nuclei (AGN) \citep{mush00}. \citet{giacconi01} resolved 60\%-80\% of the hard XRB in the \emph{Chandra} Deep Field South, and found that the X-ray and optical properties of the sources were consistent with AGN being the dominant population. Optically selected AGN have spectra characterized by strong emission lines with flux ratios that imply a powerful ionizing source, assumed to be the accretion disk around the central supermassive black hole. Surprisingly, though, more than half of the X-ray sources exist in apparently normal galaxies, with no optical indication of nuclear activity. Numerous examples of such objects have been reported in optical follow-ups of X-ray surveys \citep{fiore00, horn01, barger01, comastri02}. In the aforementioned \emph{Chandra} study, \citet{giacconi01} conclude that several of their hard XRB sources are likely to be obscured AGN, due to their red colors, and point out one object in particular that appears to be a normal elliptical galaxy.  In fact, 40\% - 60\% of all X-ray selected AGN lack optical signatures of nuclear accretion \citep{moran02}. There are two main classes of such objects: optically-elusive AGN, which have emission lines but are classified as HII regions or star forming galaxies instead of AGN by traditional line ratio diagnostics, and X-ray bright optically normal galaxies (XBONGs), which have no optical indication of nuclear activity at all, and whose optical spectra resemble quiescent normal galaxies.

Much effort has been devoted to studying these anomalous objects. Interpretations are broad and contradictory. One of the earlier objects discovered by \citet{fiore00} was further analyzed by \citet{comastri02} and found to be a Compton-thick, fully obscured AGN with nuclear X-ray emission that has been scattered and reprocessed. However, \citet{cacc07} find that none of the elusive AGN in the XMM-Newton Bright Serendipitous Survey (XBS) are Compton-thick.

Another possibility is the dilution of the spectrum due to high inclusion of starlight in the slit during observation. Indeed, \citet{cacc07} invokes this explanation for the majority of low-luminosity optically elusive AGN in their sample. For the high luminosity sources, the authors blame Compton-thin absorption for the lack of optical lines.  \citet{moran02} investigate this possibility by integrating the host galaxy light of several nearby known Seyfert 2 galaxies and adding it to the observed AGN spectrum. The authors find that, while some evidence of optical lines still exists in the integrated spectra, the objects would not have been selected as AGN in normal signal-to-noise spectra. This demonstrated that starlight dilution can cause optical normalcy in AGN. However, \citet{cocchia07} find in the HELLAS2XMM survey that while XBONGs are similar to traditional narrow-line AGN in terms of X-ray luminosity, optical luminosity, and absorbing column densities, they have systematically lower \oiii $\lambda5007$ emission. Such a finding argues against XBONGs simply being normal AGN occupying the low end of the $L_{opt} / L_X$ distribution.

More exotic explanations include radiatively-inefficient accretion flows (RIAFs), examined by \citet{yuan04} and found to be consistent with the spectra of two XBONGs, massive absorption by widespread dust in the host galaxy \citep{rigby06}, and BL Lac objects with unusually low radio luminosity \citep{brusa03}. The first two of these will be examined in depth later in this paper. The third explanation is unlikely, due to the low space-density of BL Lac objects and the relatively high space-density of XBONGs. 

If dust obscuration is a key player in the optical normalcy of these objects, examining them in the infrared is perhaps the most promising way to find the hidden nuclear emission, since extinction in the K-band is less significant than in the optical. Here, we present near-infrared (NIR) spectra of four XBONGs and four optically elusive AGN at low redshifts, to determine whether they can be successfully detected as AGN in the NIR waveband. In Section~\ref{sec:selection}, we explain how we chose our eight objects. Section~\ref{sec:reduction} discusses our observations and data reduction process. In Section~\ref{sec:agnsig} we discuss methods of infrared AGN selection and whether or not each method is successful for objects in our sample. Section~\ref{sec:dust} discusses dust obscuration, the most obvious explanation for optical normalcy that warrants investigation, and whether or not this is a likely explanation for our objects. Section~\ref{contfits} describes the results of our NIR continuum fitting.  Section~\ref{sec:alt} examines alternative explanations for optical normalcy, including starlight dilution, nuclear star formation, and radiatively inefficient accretion flows. We conclude, in an appendix, with a summary of our results for each of our eight objects and which scenario we believe to be most plausible in each case. 

\section{Sample Selection}
\label{sec:selection}
The Burst Alert Telescope (BAT) hard X-ray detector is a component of the \emph{Swift} gamma-ray burst observatory. It is sensitive to the ultra-hard X-ray band, from 14 - 195 keV, and detects X-ray sources down to a level of 4.8$\sigma$. The \emph{Swift}-BAT survey is an all-sky survey of the ultra hard X-ray band \citep{baum13}. The flux limit of the survey is $1.03\times10^{-11}$ erg cm$^{-2}$ s$^{-1}$  over 50\% of the sky, and $1.34\times10^{-11}$ erg cm$^{-2}$ s$^{-1}$ over 90\% of the sky.

To identify possible XBONGs or optically elusive AGN, we used \emph{Swift}-BAT AGN from the 58 month catalog that are nearby ($z<0.075$) with available SDSS or 6DF spectroscopy in the northern hemisphere ($\delta > -25$).  This included 94 spectra.  Of these we found eight cases with optical emission features consistent with XBONGS or optically elusive AGN. The galaxies and their classifications are listed in Table~\ref{t:tab1}. 

\section{Observations and Data Reduction}
\label{sec:reduction}

We observed our eight objects using the short wavelength cross-dispersed mode (SXD) of the SpeX instrument at the NASA Infrared Telescope Facility (IRTF).  The instrument used in this mode covers a wavelength band of 0.8 - 2.4 microns and is described in detail in \citet{rayner03}. For all observations, a $0.8'' \times 15''$ slit was used. The galaxies were observed using a nodding ABBA source pattern. The IRTF has a chopping secondary mirror for rapid sky sampling, producing AB pairs in the data. Standard A0 V stars with similar airmasses were observed after each galaxy, to provide a benchmark for telluric correction. 

The spectra were extracted using SPEXTOOL\footnote{SPEXTOOL is available from the IRTF web site at http://irtf.ifa.hawaii.edu/Facility/spex/spex.html}, a software package developed especially for IRTF SpeX observations \citep{cushing04}. Each AB pair was extracted separately, then combined using a robust weighted mean to produce an average image and scaled using the order with the best signal-to-noise. 
The combined spectra were tellurically corrected using the method described by \citet{vacca03} and embodied in the IDL routine XTELLCOR, using a Vega model modified by deconvolution. The orders were then merged into a single spectrum, and the routine XCLEANSPEC was used to remove regions of the spectrum that were completely absorbed by the atmosphere. The spectra were then smoothed using a Savitzky-Golay routine, which preserves the average resolving power.

Six of our objects are spectroscopic targets in the Sloan Digital Sky Survey (SDSS) \citep{york00}. One object, NGC 4686, has SDSS photometry but no spectrum. For this object, we have obtained an optical spectrum from Kitt Peak Observatory. The object 2MFGC 00829 is not included in SDSS at all, and is instead in the 6DF survey of the southern sky \citep{jones04}. To supplement the 6DF spectrum, we have obtained a higher S/N optical spectrum with the SNIFS instrument on the University of Hawaii 88 inch telescope. We also make use of the public data from the Two Micron All Sky Survey (2MASS) and Wide-field Infrared Survey Explorer (WISE), as well as drawing data from the literature on the \emph{Swift}-BAT survey and published data from the \emph{Spitzer} mid-to-far infrared telescope. 

\begin{figure}[ht]
\begin{center}
\plotone{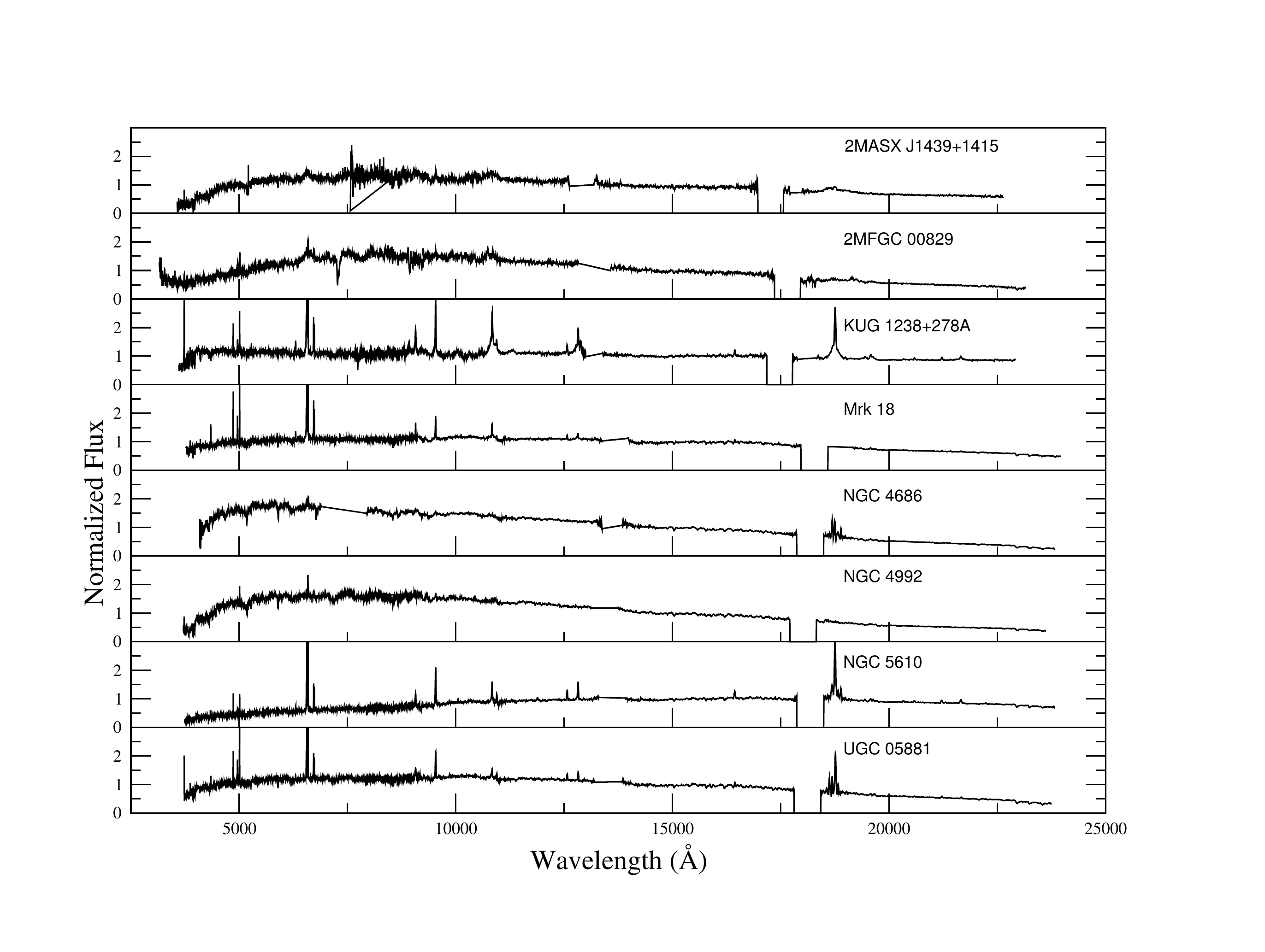}
\figcaption[]{Combined optical and infrared spectra for the four XBONGs (2MASX~J1439+1415, 2MFGC~00829, NGC~4686, NGC~4992) and four optically elusive AGN (KUG~1238+278A, Mrk~18, NGC~5610, UGC~05881). The flux has been normalized to the flux at 15000 \AA, and axes have been scaled to maximally demonstrate the respective slopes. XBONGs are typically featureless and do not exhibit emission lines, while optically elusive AGN do exhibit emission lines, but are misclassified spectroscopically as star-forming HII regions. One object, KUG~1238+278A, has a hidden broad line region revealed in the NIR.  
\label{fig:spectra}}
\end{center}
\end{figure}

The combined optical and infrared spectra are shown in Figure \ref{fig:spectra}, where the optical spectra were obtained from the various sources described above. 

\section{AGN Signatures in the Infrared}
\label{sec:agnsig}

Since the X-ray is an expensive and technically challenging band in which to observe, it is desirable to find unambiguous signatures of AGN activity in the substantially cheaper NIR band, which can be easily observed from the ground. Additionally, the extinction in the K-band is reduced by a factor of ten from the extinction in the optical, potentially offering a less-obscured view of the nuclear emission. Indeed, one of our objects, KUG~1238+278A, exhibits broad emission lines in the NIR spectrum, visible in Figure \ref{fig:spectra}. In this case at least, we can be certain we are seeing the nucleus through a screen of dust that obscured AGN activity in the optical spectrum. 

We provide here a summary of infrared AGN selection techniques and whether they recovered objects in our sample. The results of each test for each object are summarized in Table \ref{t:tab2}. In short, we find that NIR emission line ratio diagnostics are successful in classifying elusive AGN where optical ratios fail, but that these ratios do not differentiate control samples cleanly. We find that coronal emission lines indicative of the hard ionizing continuum of AGN are not present in the NIR spectra of our sample, but that we do detect coronal lines in the mid-IR (MIR) for the subset of our objects with \emph{Spitzer} spectra.  We also conclude that infrared photometric techniques are unsuccessful in classifying the majority of our objects. After we have discussed whether these diagnostics recover our optically normal objects, we proceed to considering the evidence of possible reasons for their optical normalcy.
\subsection{Line Ratios: Optical and Infrared}
\label{sec:linerat}

Optical selection of AGN is typically done in two ways. Type~1~AGN are selected by searching for broad emission lines, usually H$\alpha$, H$\beta$, and \mgii~$\lambda~2798$\AA. These objects have broad emission lines because our viewing angle allows us to see the gas moving in the gravitational potential of the supermassive black hole. The area where the gas that produces broad emission lines resides is called the broad-line region (BLR). These lines are typically Doppler broadened to widths greater than 1000 km/s. Type~2~AGN are viewed through the obscuring dusty torus surrounding the central engine. Consequently, we can only see the narrow emission lines, from gas ionized by the accretion disk but located above the vertical extent of the dusty torus. The region from which this gas is emitted is referred to as the narrow-line region (NLR). Since star formation produces many of the same emission lines as gas ionized by the central accretion disk, one must use line ratios to distinguish the source of the ionizing radiation. The accretion disk has a significantly harder ionizing continuum than HII regions can produce, which manifests in different emission line ratios.

\begin{figure}[t!]
\begin{center}
\plotone{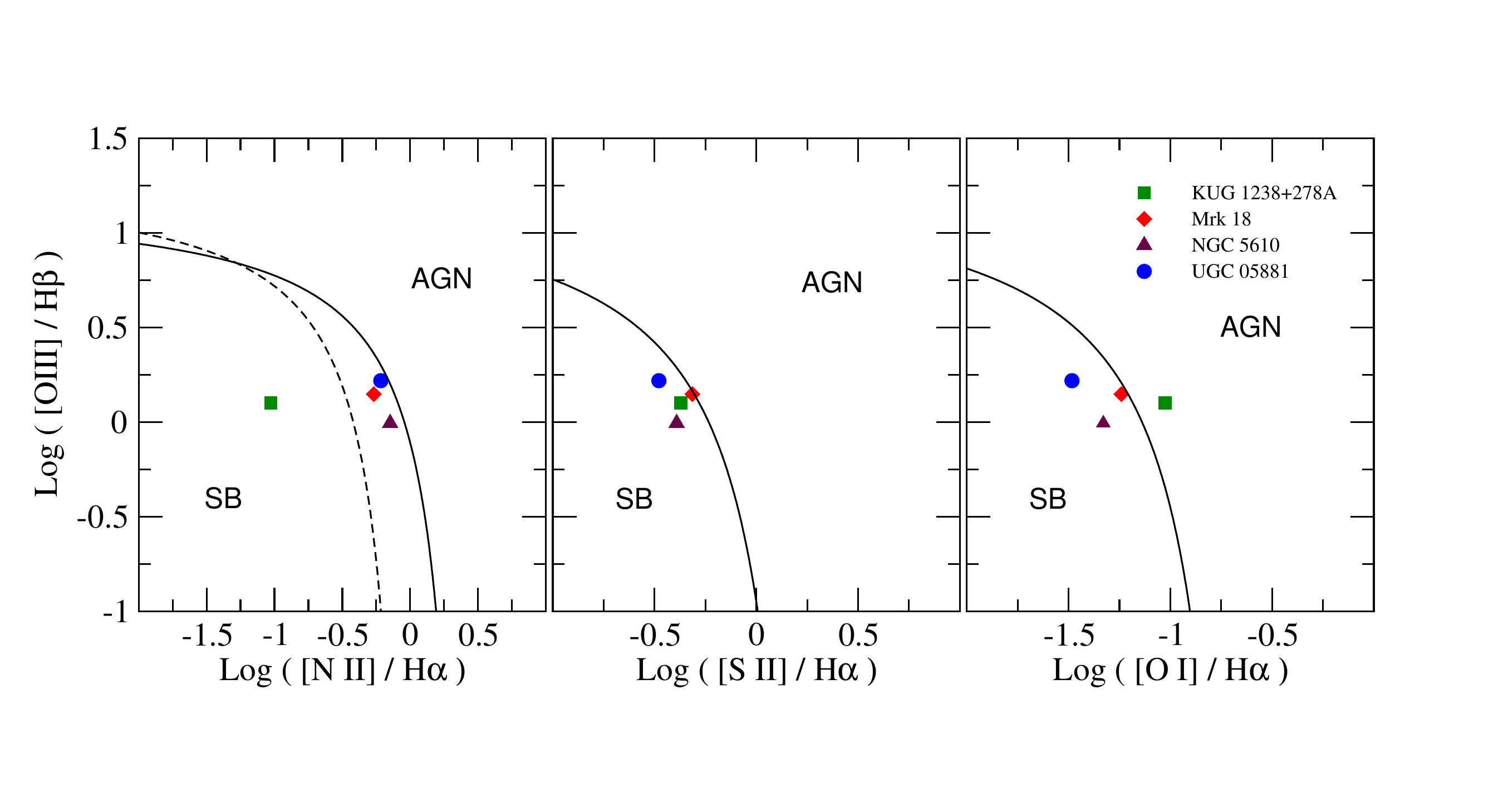}
\figcaption[]{Our four optically elusive AGN plotted on the optical diagnostic diagrams of \citet{kewley06}. None of the XBONGs exhibited enough of the lines to construct the ratios. The dashed line in the first panel indicates the pure star-formation line according to \citet{kauffmann03}. All of the objects fall in the HII region portion of the Kewley diagrams,  despite being unambiguous AGN in the ultra-hard X-ray detection of \emph{Swift}-BAT, except for KUG 1238+278A which falls in the AGN portion of the [OI]/H$\alpha$ diagram.
\label{fig:kewley}}
\end{center}
\end{figure}

Four of our eight sources are categorized as optically-elusive, meaning that they exhibit emission lines, but that the traditional line ratio diagnostic BPT (Baldwin-Phillips-Terlevich) diagram \citep{baldwin81} characterizes them as star-forming HII regions instead of AGN. \citet{veilleux87} first published the revised version of this diagnostic plot, taking advantage of three sets of line ratios and excluding ratios badly affected by reddening. \citet{kewley06} published the modern version of this diagram.  Using the spectral fitting routine \texttt{GANDALF} (for more on \texttt{GANDALF} see Section \ref{sec:nucdust}), we have calculated the fluxes in the lines necessary for these diagnostics and provided the values in Table~\ref{t:tab3}. For XBONGs with no evidence of lines, we have provided the 2-$\sigma$ flux upper limits. Figure~\ref{fig:kewley} shows the optical diagnostics discussed above for the four optically elusive objects. None of the XBONGs have at least one detected line in both ratios, so we cannot plot upper limits. The curve in these three figures delineates star-forming regions from AGN. Objects in the lower left of the diagram, below the curve, are classified as HII regions associated with star formation. All four of our objects fail to be classified as AGN by these traditional means, despite being unambiguous AGN in the ultra-hard X-ray. 

Since infrared emission is less sensitive to dust extinction than optical, it is possible that such optically elusive objects might be identifiable as AGN by near infrared diagnostics instead. If so, the NIR is a cheap alternative to X-ray selection for building a more complete AGN sample which includes obscured objects. \citet{rodrig04} undertook an in-depth study of the excitation sources for the strongest NIR spectral features of $H_2$ and \feii~in various emission line galaxies. In general, the vibrational and rotational modes of $H_2$ can be excited by three mechanisms: UV fluorescence, shock heating, and X-ray heating. Each of these mechanisms excites the various $H_2$ modes differently, so it is possible to tell which is the dominant mechanism by comparing ratios of different $H_2$ lines. \citet{rodrig04} perform this analysis and find that, in AGN, the $H_2$ excitation is thermal, with only a 15\% contribution from non-thermal excitation. This precludes a significant UV fluorescence contribution to the observed $H_2$ in AGN, but it does not distinguish between shocks from the central radio jet or X-ray illumination.

\feii~emission in the NIR can also be used to distinguish between dominant emission-line excitation mechanisms. A strong correlation between the \feii~emission and the 6 cm radio emission in the central regions of AGN, reported by \citet{forbes93}, suggests that \feii~in Seyfert galaxies is predominantly produced by shocks from the central radio jets. \citet{simpson96} and other studies have shown that shocks from supernova remnants can produce \feii~in starburst galaxies. The amount of \feii~produced by photoionization in HII regions versus shock excitation can be constrained by comparing ratios of the \feii~emission with Pa~$\beta$. The ratio of \feii~to the hydrogen lines will increase as you move from pure photoionization to pure shock excitation. \citet{alonso97} first found that Seyfert galaxies typically have ratios 0.5~$<$~\feii~/~Pa~$\beta~<$~4.6, and \citet{rodrig04} confirm this finding.  So, \feii~/~Pa~$\beta$ can be used to determine whether the \feii~emission is excited by stellar or nonstellar sources. The nonstellar sources are either X-ray heating or shocks from radio jets; both are phenomena associated with AGN.

\citet{larkin98} were the first to combine both H$_2$~2.12~$\mu$m and [Fe~{\sc ii}]~1.257~$\mu$m into a diagnostic of nuclear activity,  noting that $H_2$~/~Br~$\gamma$ and \feii~/~Pa~$\beta$ are linearly correlated, with starburst galaxies occupying the lower end of the relation and AGN occupying the intermediate range. Low Ionization Nuclear Emission-line Region galaxies (LINERs) occupy the highest end of the relation.

Unfortunately, the four XBONGs in our sample do not even have lines in the NIR (NGC~4992, for example, has a featureless continuum across the entire displayed SED; see Figure~\ref{fig:spectra}), so we cannot carry out this test for them. The optically elusive objects, however, do exhibit all the necessary emission lines to place them on the IR diagnostic diagram first put forth by \citet{larkin98} and refined by \citet{rod05}. The most recent version of this diagram belongs to \citet{riffel13}, who enlarged the sample and included supernova remnants and blue compact dwarf galaxies, which are comprised mainly of star forming regions. Their improved limits on line ratios for various levels of activity are as follows: for starbursts,  \feii / Pa~$\beta \lesssim 0.6$ and $H_2$ / Br~$\gamma \lesssim 0.4$; for AGN, $0.6 \lesssim$  \feii / Pa~$\beta \lesssim 2$ and $0.4 \lesssim$ $H_2$~/~Br~$\gamma \lesssim 6$; and for LINERs, \feii~/~Pa~$\beta \gtrsim~2$ and $H_2$~/~Br~$\gamma \gtrsim~6$.

\begin{figure}[ht]
\begin{center}
\plotone{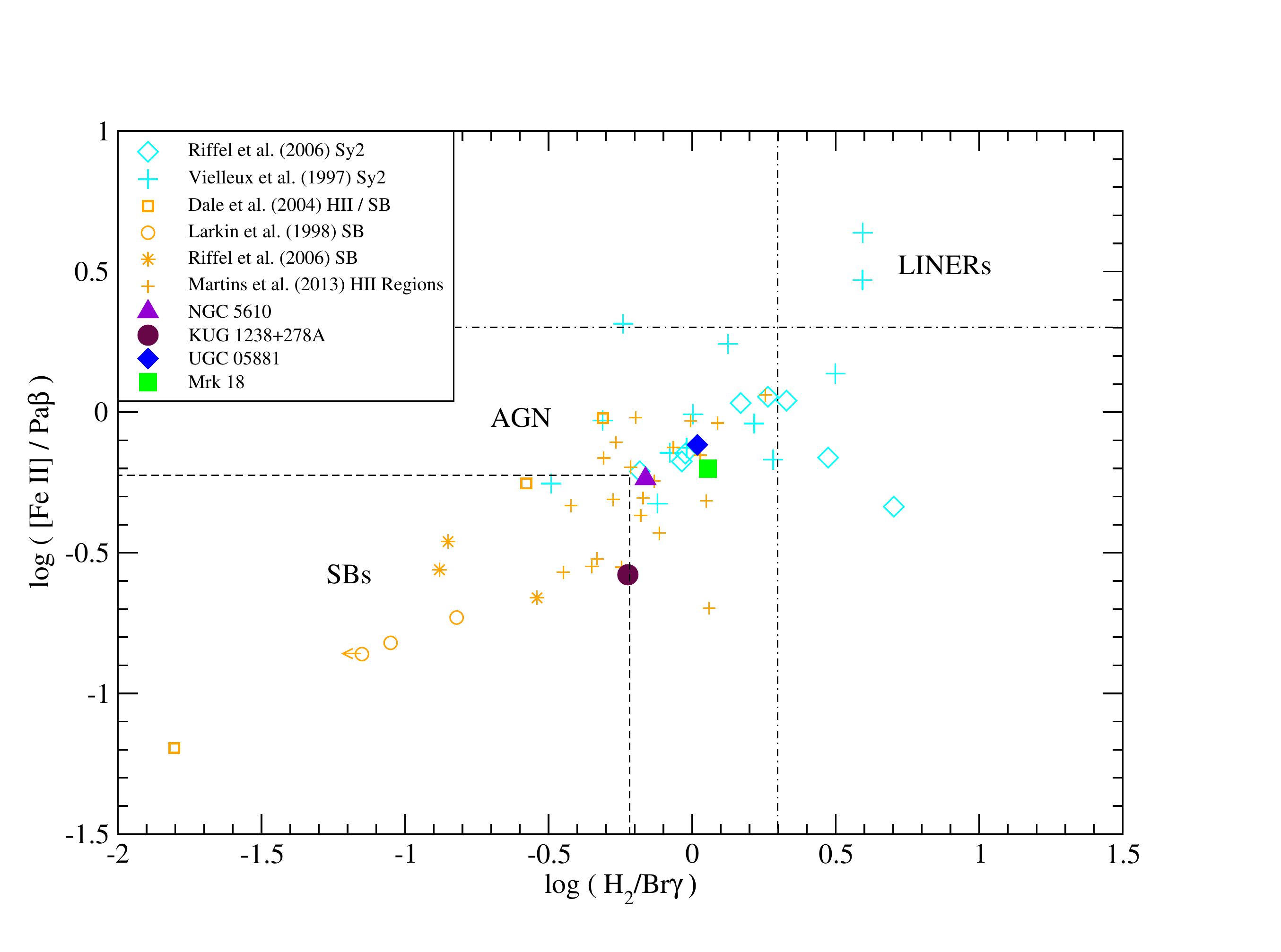}
\figcaption[]{Infrared emission line diagnostics from \citet{rodrig04} and \citet{riffel13}. Plotted are the four optically elusive AGN which have measurable emission lines in the NIR, the sample of starburst galaxies and Type 2 AGN from \citet{riffel06}, starburst and HII region samples from \citet{dale04}, \citet{larkin98}, and \citet{martins13}, and Type 2 AGN from \citet{v97}. While our objects do have values typical of AGN, the diagnostic diagram is not able to distinguish between AGN and HII regions/SBs cleanly.
\label{fig:irdiag}}
\end{center}
\end{figure}

We measured the flux in the H$_2$ 2.12~$\mu$m, Br~$\gamma$, [Fe~{\sc ii}]~1.257 $\mu$m, and Pa~$\beta$ lines using the IRAF routine SPLOT. The values are given in Table~\ref{t:tab3}, and ratios are plotted in Figure~\ref{fig:irdiag}.  All four objects occupy the region of the plot which, according to the references above, typifies AGN. Also plotted are the samples of starburst galaxies and Type~2~AGN from \citet{riffel06}, starburst galaxies and HII regions from \citet{dale04}, a sample from the \citet{martins13} NIR atlas of HII regions, and a sample of Type~2~AGN from \citet{v97}. 
As both \citet{riffel13} and \citet{martins13} acknowledge, while AGN are indeed located in the region with intermediate values of both line ratios, starbursts and HII regions can have both low and high values on both axes and encroach quite far into the AGN regime. 

Our four optically-elusive objects fall in the AGN region of the NIR diagnostic diagram, despite failing optical tests. While this is promising evidence that the weaker extinction in the NIR allows us to spectroscopically detect nuclear activity, since the diagnostics fail to clearly distinguish between star forming regions and AGN, we do not consider this a successful method of recovering buried AGN. There remains no way, by NIR line ratios alone, to determine whether the source is an active nucleus or a starburst. 

\subsection{Coronal Emission Lines in the NIR}
\label{coronal}

It has been mentioned that UV radiation from star forming regions can excite a similar emission spectrum to that excited by nuclear activity. While this is true, the hard ionizing photons from an AGN accretion disk are capable of exciting much higher ionization lines than an HII region. The coronal lines are collisionally-excited forbidden transitions in highly ionized species, usually with ionization potentials greater than 100 eV. Such IPs require much harder ionizing radiation than HII regions can emit, and so are considered definite indicators of AGN activity. Several coronal lines exist in the region covered by our NIR spectra; namely, [Si~{\sc x}] 1.43$\mu$m, [Si~{\sc vi}] 1.963$\mu$m, and [Ca~{\sc viii}] 2.32$\mu$m. We searched for these three lines in our NIR spectra, and were unable to confirm the detection of coronal lines in any of the XBONGs or optically elusive AGN.  It is important to note that one would not necessarily expect to find these coronal lines in all Type 2 AGN; \citet{rodrig08} report that of a sample of classical Type 2 AGN, [Si~{\sc vi}] was detected in a little more than 40\% of the spectra, while [Si~{\sc x}] was detected in less than 25\%. They report that in general, high-ionization coronal lines appear less frequently in Type 2 than in Type 1 objects. All of our optical spectra would be considered Type 2, since they do not exhibit broad optical emission lines. Therefore, the lack of coronal line detection does not imply the lack of an AGN, but their detection does imply the presence of an AGN.

\subsection{Mid-Infrared Emission Lines}
\label{nev}

\begin{figure}[ht]
\begin{center}
\plotone{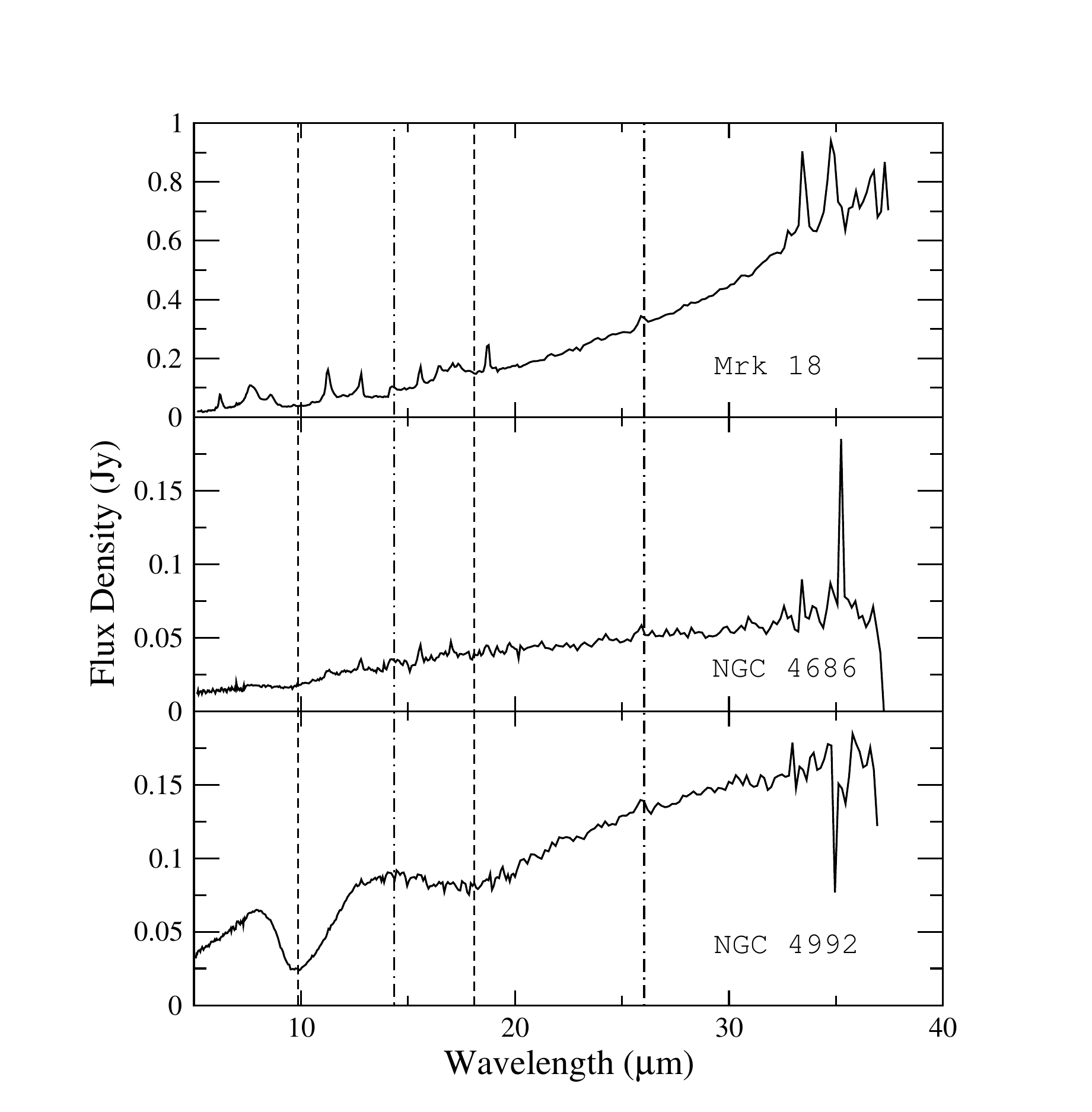}
\figcaption[]{Mid-infrared spectra for the three objects in the \emph{Spitzer} IRS archives. The dashed lines indicate the positions of the 9.7 $\mu$m and 18 $\mu$m silicate absorption bands. The dash-dot lines indicate the positions of the \nev~$\lambda$14.32~$\mu$m and the \oiv~$\lambda$25.90~$\mu$m high-excitation lines. Mrk 18 is an optically elusive AGN, while NGC 4686 and NGC 4992 are XBONGs. The silicate absorption feature, indicative of dust along the line of sight, is prominent in NGC 4992 and detected in NGC 4686, but absent in Mrk 18. Star-formation related PAH emission is prominent in Mrk 18. 
\label{fig:spitzer}}
\end{center}
\end{figure}

Three of our objects have public data from the \emph{Spitzer} mid-to-far IR telescope \citep{weaver10}. The \nev~line at 14.32$\mu$m and the \oiv~line at 25.90$\mu$m, like the coronal lines mentioned above, have ionization potentials requiring an AGN accretion disk as the ionizing source. The \emph{Spitzer} spectra for these three sources are shown in Figure \ref{fig:spitzer}. One source, Mrk 18, is an optically elusive AGN; the other two (NGC 4686 and NGC 4992) are XBONGs. For \nev, we have a clear detection in Mrk~18, a possible but not significant detection in NGC~4686, and no detection in NGC~4992. All three objects exhibit \oiv~emission, even the heavily obscured NGC~4992.

There have been AGN diagnostic techniques developed using MIR emission line ratios. \citet{genzel98} use the ratio \oiv$\lambda25.9~\mu$m / \neii$\lambda12.8~\mu$m with the strength of the 7.7$\mu$m PAH feature to identify the source of the nuclear emission in \emph{IRAS} galaxies. They find that AGN-dominated sources typically have weak 7.7$\mu$m PAH emission and a high \oiv / \neii~ratio. Neither NGC~4686 nor NGC~4992 have the requisite PAH feature, but Mrk~18 has all of the necessary lines to carry out this diagnostic. For Mrk~18, we find that \oiv~/~\neii = 0.28$\pm$0.08 and the relative strength of the 7.7$\mu$m PAH feature to be 2.92$\pm0.12$ (\citealt{genzel98} define this quantity as the ratio of the peak 7.7$\mu$m flux to the 7.7$\mu$m continuum flux as measured by a linear interpolation between 5.9 and 11.0~$\mu$m). These numbers place Mrk~18 solidly in the region of their diagram populated by objects with approximately equal contributions of AGN and star formation.

\subsection{Infrared Photometric Selection Techniques}
\label{irphot}

AGN are among the very few astronomical objects with strong emission throughout the entire infrared spectral range. Typically, AGN continua in this regime are power-laws; in particular, the MIR shape of AGN continua is best fit by a function $f_\nu \propto \nu^{\alpha}$, with $\alpha \leq -0.5$ \citep{alonso06}. This has motivated many investigations into AGN selection by mid-IR photometry using the Wide-Field Infrared Survey Explorer (WISE), which surveyed the entire sky in four photometric bands (3.4, 4.6, 12, and 22 $\mu$m). Photometric accuracy in the 22 $\mu$m band is not typically reliable and coverage in that band is significantly shallower than the other three, so only the first three bands are used in most AGN selection schemes. 

Selection of AGN by MIR colors should be more complete than in the X-ray, since even in highly obscured AGN, the nuclear emission is re-radiated by dust in the MIR. However, if severe dust extinction affects the flux at the lower WISE wavelengths, and if the MIR colors are not dominated by the AGN thermal continuum but instead by the starlight from the host galaxy, an AGN will fail the photometric power law criteria.

All eight of our objects have WISE data. To see whether they would be classified as AGN based on MIR photometry, we use three different WISE selection schemes from the literature. First, we use the classification from \citet{stern12}: a color cutoff of [3.4]-[4.6] $\geq~0.8$, which produces an AGN sample with 98\% reliability and 78\% completeness. Second, we use the three-band  ``AGN-wedge" defined by \citet{mateos12}, the completeness of which is strongly dependent on AGN luminosity. Low-luminosity AGN are more likely to have MIR colors not defined by the thermal emission from the AGN, and so more frequently fail their test. In particular, only $\sim$39\% of Type 2 AGN with $L_X<10^{44}$~erg~s$^{-1}$ fall inside their AGN color wedge. At $L_X>10^{44}$~erg~s$^{-1}$, $\sim$76\% of Type 2 AGN are inside the wedge. All but one of our objects, 2MASX~J1439+1415, have X-ray luminosities in their lower range. Finally, we use the $S_I$ algorithm defined by \citet{edelson12}, which uses the first three \emph{WISE} bandpasses and finds that most objects with $S_I<0.888$ were non-AGN while most with $S_I>0.888$ tended to be AGN. A summary of which objects pass and fail these NIR photometric tests is given in Table~\ref{t:tab2}. 

\begin{figure}[ht]
\begin{center}
\plotone{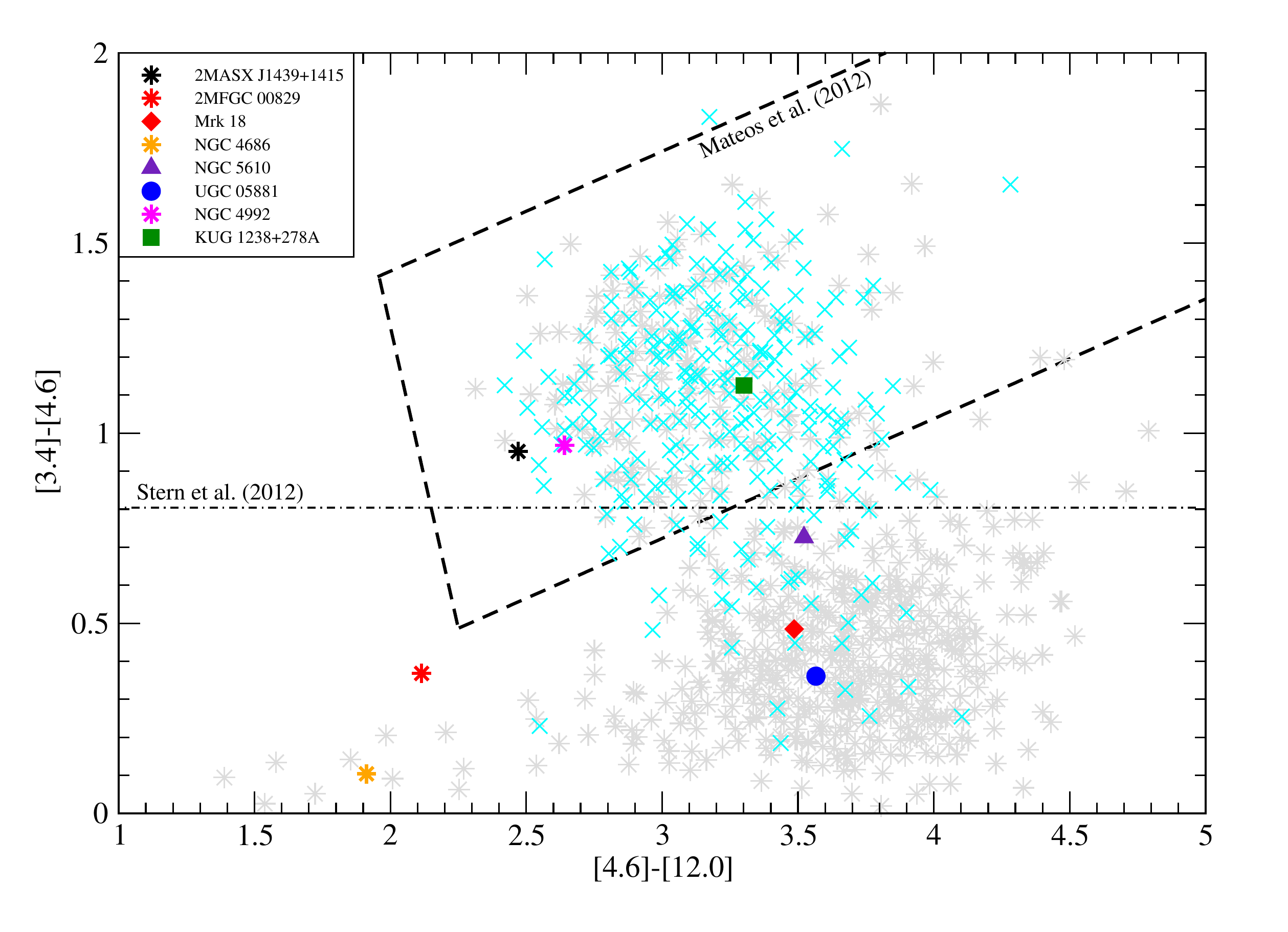}
\figcaption[]{\emph{WISE} three-color diagram with \citet{stern12} and \citet{mateos12} AGN selection criteria. For comparison we have also plotted the control samples from \citet{mateos12}: \emph{WISE} sources with X-ray detections (cyan crosses) and without X-ray detections (grey stars). The XBONGs are denoted as stars, while the optically elusive objects have the same solid symbols as in the other figures. Most of our objects fail both selection criteria; those that pass are the hidden BLR object (KUG 1238+278A), our best candidate for starlight dilution (2MASX J1439+1415), and, surprisingly, the XBONG NGC 4992. 
\label{fig:irphot}}
\end{center}
\end{figure}

Figure \ref{fig:irphot} shows how our objects are classified based on \emph{WISE} colors in the \citet{stern12} and \citet{mateos12} schemes. Most of the objects conform to our expectations for these tests. Of the optically elusive AGN, KUG~1238+278A is the only one that passes the AGN criteria. This happens for the same reason that we see broad lines in the NIR spectrum: the dust extinction is minor enough that we can peer through it in the NIR, revealing the nuclear spectrum. The other optically elusive AGN fall outside the selection regions, in the cloud of \emph{WISE} sources without X-ray detections (grey in Figure \ref{fig:irphot}). There are several outliers, however, of the sample of \emph{WISE} sources with X-ray detections (cyan in Figure \ref{fig:irphot}) which occupy that same region; we assume, then, that our optically elusive \emph{Swift}-BAT AGN are among them. 

For reasons discussed in detail in later sections, we believe the XBONG 2MASX~J1439+1415 is a heavily starlight-diluted AGN that would otherwise be an optical Type 1 and is not subject to major dust obscuration. Additionally, it is the only object in the sample with $L_X>10^{44}$~erg~s$^{-1}$ and has a high Eddington ratio, unlike the rest of our sample. The object passes all three photometric selection criteria. For more detailed discussion of this object, see Sections \ref{sec:dilut} and Section~\ref{sec:2mas}. 

More complex is the case of NGC~4992. This XBONG is heavily dust obscured and has a NIR continuum shape that is definitely not defined by the AGN thermal continuum (see Section~\ref{contfits}). We successfully detect \oiv~in this object's MIR \emph{Spitzer} spectrum. Perhaps the combination of a strong but obscured nuclear component (assumed from the high X-ray luminosity and column) and the fact that we begin to penetrate this thick dust in the MIR (as judged from the \oiv~detection) leads to NGC~4992 passing these criteria. NGC~4992 is discussed in detail in Section~\ref{sec:4992}.
 
\section{Dust Extinction}
\label{sec:dust}

The most plausible explanation for the lack of nuclear activity detection at optical wavelengths is extinction by circumnuclear or host galaxy dust. An optically thick dust distribution concealing the central engine could cause the optical spectrum of a true AGN to resemble only the host galaxy, and the emission lines to reflect the star formation in the host that would otherwise be washed out by light from the nucleus. Several tests for significant dust extinction are possible in the optical and NIR. One can look for visible dust in the images, if the obscuring dust is thought to reside in the host galaxy at large. Additional tests include calculation of the $E(B-V)$ reddening, measuring the X-ray column densities (shown in Table \ref{t:tab1}), and looking for silicate absorption features in the MIR. 

As a first handle on the obscuration, we have measured the X-ray absorbing column density along the line of sight from the photoelectric cutoff in the X-ray spectrum, if the S/N of that spectrum is adequate for the task. All of our objects except for 2MFGC 00829 have \emph{Swift} XRT spectra. Only KUG 1238+278A has an XRT spectrum with too much noise to confidently assess the column density. The calculated values of $N_H$ are given in Table \ref{t:tab1}. As in \citet{cacc07}, none of our optically normal AGN are Compton-thick ($N_{H}~\gtrsim~10^{24}$~cm$^{2}$).

\subsection{Dust in the Host Galaxy}
\label{sec:hostdust}

\citet{rigby06} posit that widespread dust in the host galaxy could be responsible for XBONGs. For a sample of only eight objects, a statistical examination of the axis ratios would not be enlightening. We do, however, have good images for most of the objects. 

The optical images for our eight objects are shown in Figure~\ref{fig:images}. All but 2MFGC 00829 were obtained from SDSS; that object is out of the Sloan footprint and so we present here the 2MASS JHK image. Only one of our objects, NGC 4686, shows a prominent dust lane potentially in the line of sight. Since 2MFGC 00829 is apparently an edge-on spiral, host galaxy dust is possible in this object as well. The other galaxies appear to be face-on or do not have obvious dust lanes, and 2MASX~J1439+1415 is elliptical, and so are unlikely to be plagued by large-scale host galaxy dust. Only one object, KUG 1238+278A, appears to be interacting.

Absorption in the host galaxy beyond the torus was considered by \citet{civano07}, but these authors found no evidence of large-scale dust structure in HST images of two of their XBONGs, similar to our result.

While we do not see obvious widespread dust in the host galaxies of our objects, a simple subjective analysis of the images is not enough to rule out host galaxy dust as the responsible extinctor.

\subsection{Circumnuclear Dust}
\label{sec:nucdust}

Since the dust does not appear to be present in the host at large, it becomes necessary to check for the existence of circumnuclear dust which could screen the gas ionized by the central engine. To perform a robust calculation of the dust reddening affecting the emission lines, we use the spectrum fitting code \texttt{GANDALF} \footnote{\texttt{GANDALF} was developed by the SAURON team and is available from the SAURON website (www.strw.leidenuniv.nl/sauron). See also \citet{sarzi06} for details.}. The code first fits a stellar continuum while masking the regions of the spectrum that contain emission lines, and then measures the gas emission on the residual spectrum of the stellar fit. In order to estimate the intrinsic reddening of the source, the program adopts the dust model of \citet{calzetti00}. 

Examples of \texttt{GANDALF} fits are shown in Figure~\ref{fig:gandalf}.

The calculated $E(B-V)$ values are listed in Table \ref{t:tab1}. We exclude values for the XBONGs since the \texttt{GANDALF} code was only able to find upper limits on the fluxes of the Balmer lines in those objects. As a control sample, we used the OSSY\footnote{The OSSY database is a collection of quality-assessed emission and absorption line measurements of all galaxies in the SDSS DR7, including stellar and gaseous kinematics, and is available at http://gem.yonsei.ac.kr/$\sim$ksoh/wordpress/. } database of all SDSS DR7 galaxies with $z<0.2$ which have been run through the \texttt{GANDALF} program \citep{oh11}. We obtain a control sample average of $E(B-V)=0.015\pm0.0002$. So, the optically elusive AGN have an order of magnitude higher dust extinction than the control sample on average. One should have caution, though: while the average reddening for the control sample is very robust due to its large size, our sample is small. 

With the $E(B-V)$ values in hand, we may convert them to extinction values via the standard formula $A_{V} = 3.2 \times E(B-V)$, and use the well-studied relationship between extinction and hydrogen column density as published by \citet{guver09}: $N_{H}$ (cm$^{-2}$) = ($2.21 \pm 0.09$) $\times 10^{21} A_{V}$ (mag). We find that this method predicts values of $N_{H}$ about one order of magnitude, on average, below the values measured from the X-ray spectra in our objects. This is not unusual, and was first remarked upon by \citet{macca82}. \citet{maiolino01} provide observational confirmation that dust in the region surrounding the nucleus is not homologous with dust in the ISM of the host at large. Particularly relevant for our objects, those authors find that the $A_{V} / N_{H}$ ratio is significantly lower than the Galactic standard value in many AGN categories, including hard X-ray selected AGN. This remains true for our sample.

\subsection{Morphology: Bars and Mergers}
\label{morph}

One explanation for the presence of a significant circumnuclear dust population is the driving of dust to the central regions through tidal torques in mergers, or along a stellar bar. It is interesting, then, to see whether mergers or bars are overrepresented in our sample. As a control, we assemble all the AGN from the \emph{Swift}-BAT sample that have images in the SDSS, which amounts to fifty objects. Of those fifty, we subtract the eleven edge-on galaxies, since we would not be able to identify a bar in them. We must also remove the one edge-on object from our sample, 2MFGC 00829. We do not exclude edge-on objects from the analysis for interaction, since tidal tails, warped disks or a disturbed companion could be visible in such objects.

After removal of the edge-on galaxies, we find that 13/39 control galaxies, or 33\%, have stellar bars. In our sample, 4/7, or 57\%, have bars. The small numbers prevent a reliable comparison, but it is possible that bars in our sample are contributing to circumnuclear dust and increasing our sample's obscuration compared to the other \emph{Swift}-BAT AGN. Only one of our objects, KUG 1238+278A, appears to be interacting; this constitutes 12\% of our sample. Eleven of the 50 control objects, or 22\%, are interacting. However, we refrain from drawing conclusions due to the small numbers in the sample.

The HyperLeda\footnote{HyperLeda can be reached online at http://leda.univ-lyon1.fr/ and was compiled by \citet{paturel03}.} database morphological classifications of the galaxies are given in Table~\ref{t:tab1}. 2MASX J1439+1415, KUG 1238+278A, and 2MFGC 00829 are not classified in HyperLeda. 2MASX J1439+1415 appears to be an elliptical galaxy. KUG 1238+278A is interacting, and has a peculiar morphology. 2MFGC 00829, as best we can tell from the 2MASS JHK image, is an edge-on spiral. Except for 2MASX J1439+1415, all of our objects are in spiral or interacting hosts. In general, spiral galaxies are dustier than elliptical galaxies, which may cause more XBONGs to be found in spirals due to extinction of nuclear light. However, since the BAT sample is in general dominated by spiral galaxies, it is not necessarily surprising that this remains true for our sample.

\subsection{Silicate Absorption Features}
\label{silicate}

Another signature of dust obscuring the line of sight can be found in the \emph{Spitzer} MIR data: absorption features at 9.7 $\mu$m and 18 $\mu$m due to amorphous silicate dust. \citet{shi06} find that the strength of silicate absorption features correlate with the X-ray column densities. Three of our galaxies have archival \emph{Spitzer} information (Mrk~18, NGC~4686, and NGC~4992). The \emph{Spitzer} spectra for these three galaxies are shown in Figure \ref{fig:spitzer}. Silicate absorption is apparent in both XBONGs (NGC~4992 and NGC~4686). In fact, in NGC~4992, the silicate feature totally dominates the spectrum at 9.7 $\mu$m, suggesting it is a dust-shrouded galaxy. The optically elusive AGN Mrk 18 does not exhibit pronounced silicate absorption. 

The emission features in the Mrk 18 spectrum are PAH features associated with star formation. It is possible that this object is undergoing a nuclear starburst. For further discussion, see Section~\ref{sec:sf}.

\section{NIR Continuum Fits}
\label{contfits}
We have chosen the NIR band to search for AGN signatures because extinction in this band is an order of magnitude less than in the optical. If the NIR light from these objects is truly coming from the central engine, we would expect the NIR continuum to be fit well by a power-law accretion disk and a blackbody component from the hot dust of the torus. \citet{landt11} found that the NIR continua of Type~1 AGN were well fit by this combination. However, our objects may continue to be obscured in the NIR, which we believe to be the case for all except KUG 1238+278A (our hidden broad line region object). In this case, we would expect the NIR continuum to be well fit by a stellar template, and a dust blackbody. For our rough purposes, we approximate the stellar template by a blackbody with a temperature typical of a stellar bulge population, $T_{stars}\sim3300-3700$~K. Like \citet{landt11}, we model the nuclear hot dust as a blackbody with $1100<T_{hot}<1700$, which approximates the dust sublimation temperature for a variety of grain compositions as calculated by \citet{salpeter77}. We have also included a gaussian component, to account for the broad emission feature at 1.68~$\mu$m due to a minimum in stellar atmospheric opacities \citep{john88}, typical in galaxy spectra. The fits are shown for the XBONGs in Figure~\ref{fig:xbongfit} and for the optically elusive AGN in Figure~\ref{fig:oefits}. 

All eight of the galaxies are well modeled by a combination of a stellar and hot dust blackbody. We are ignoring small features, like emission lines or absorption troughs, and attempting to fit only the generic continuum shape. A power law is a very poor fit, even when combined with a blackbody, for all of our objects; we therefore conclude that we are not seeing nuclear continuum emission, even in the NIR. A single blackbody at the stellar temperature was also a poor fit, despite the fact that the \texttt{GANDALF} fits approximated the optical continuum very well with stellar templates; thus, there is some nuclear dust contribution in the NIR. 

Insight to the relative amount of obscuration by cool dust in each object may be gained from comparing the relative contributions of the stellar and dust blackbody components. If we assume a simple model in which the the hot nuclear dust at the sublimation temperature is affected by a screen of cool obscuring dust while the stellar contribution is not, then we would expect obscured objects to have a small hot dust contribution compared to the stellar contribution.  The XBONGs  2MFGC~00829, NGC~4992 and NGC~4686 have a smaller hot dust contribution than the optically elusive objects KUG~1238+278A, Mrk 18 and NGC~5610. Via our model, we attribute this to a higher obscuration of the nuclear region of the XBONGs, including the hot dust near the sublimation temperature which is putatively on the inner edge of the dusty torus. Indeed, NGC~4686 and NGC~4992 have very high X-ray column densities, $Log(N_{H})=23.64$ and 23.69, respectively. The elusive objects, in which we presume obscuration by cool dust to be less important, we see the hot dust component kicking in sooner and contributing more significantly to the long wavelength end of the spectrum. Whether or not each object is well fit with a prominent hot dust component is summarized in Table~\ref{t:tab2}.

Interesting exceptions are the XBONG 2MASX~J1439+1415 and the elusive object UGC~05881, which exhibit opposite character from those described above. Further, UGC~05881 has the same X-ray column density as fellow elusive object Mrk 18, but has a small hot dust contribution compared to Mrk~18. The behavior of 2MASX~J1439+1415 is explained by our theory that is is a starlight-diluted Type~1~AGN. 

These fits are consistent with the hypothesis that the primary reason for optical normalcy in extreme XBONGs like NGC~4686 and NGC~4992 is heavy obscuration by a cool dust component beyond the dusty torus.  However, these results should be interpreted with caution: the relationship between the hot and cool dust populations in AGN is probably very complex, and a sophisticated analysis of the origin of each type of emission is beyond the scope of this paper.

\section{Alternative Explanations}
\label{sec:alt}

\subsection{Starlight Dilution}
\label{sec:dilut}

Follow-up optical observations of the X-ray sources from deep \emph{Chandra} surveys have shown that 40\% to 60\% of the optical counterparts are normal early-type galaxies with weak or no optical emission lines. These objects are at relatively high redshifts, and so all or most of the galaxy being observed falls inside the spectrograph slit. This results in a large amount of host galaxy starlight being integrated into the spectrum, effectively hiding the nuclear emission and resulting in a normal galaxy spectrum. 

\begin{figure}[ht]
\begin{center}
\plotone{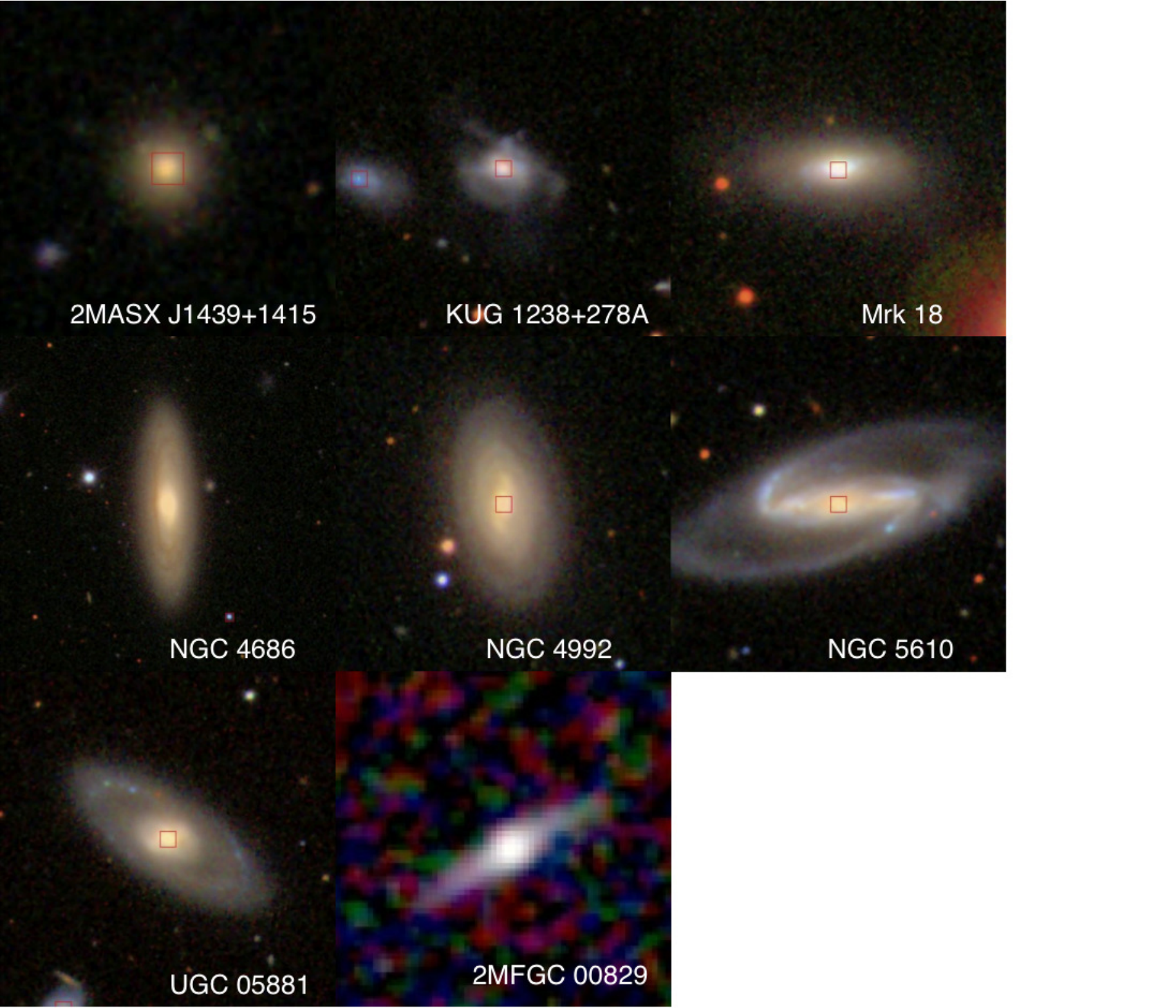}
\figcaption[]{Images for our eight XBONGs and optically elusive AGN. All of the images are optical and from the SDSS image server except for 2MFGC 00829, which is the 2MASS JHK image. For the SDSS images that have spectra in the survey (all except NGC 4686), the red box indicates the 3\arcsec~region for which the spectrum was taken. Note that another spectroscopic target exists in the same frame with KUG 1238+278A; its spectrum is not part of our sample.
\label{fig:images}}
\end{center}
\end{figure}

\begin{figure}[ht]
\begin{center}
\plotone{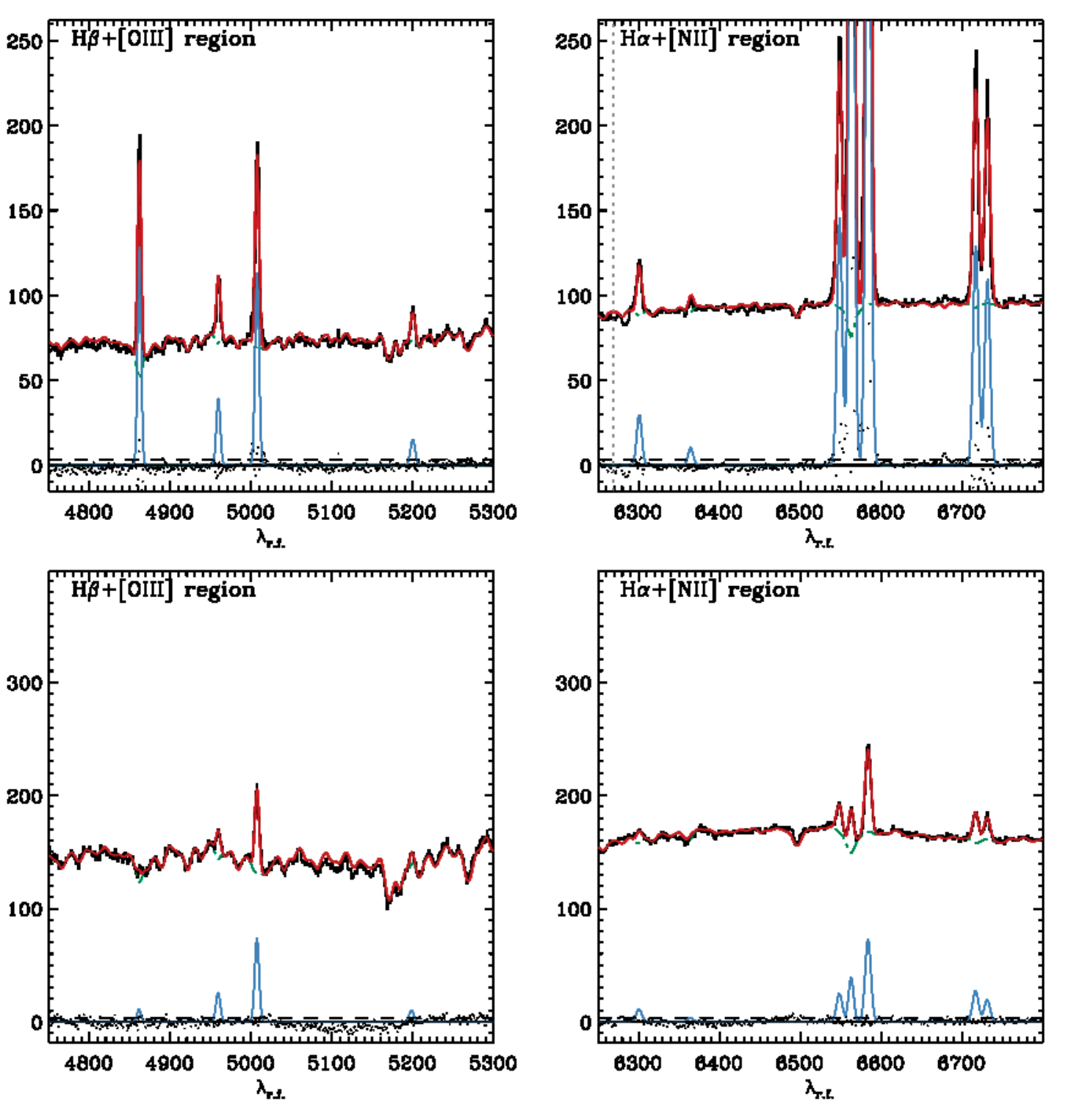}
\figcaption[]{An example of the fit that \texttt{GANDALF} performs on the spectra. The upper spectral regions are from the optically-elusive AGN NGC 5610, and the lower are from the XBONG NGC 4992. The red line is the final fit of the spectral template plus the emission lines; the lower blue lines show the emission lines only. The optical continuum is very well-fit by a stellar-only template in all of our objects (i.e., there are no residuals after subtraction of the spectral fit template). 
\label{fig:gandalf}}
\end{center}
\end{figure}

\begin{figure}
    \centering
    \subfigure{\includegraphics[width=0.45\textwidth]{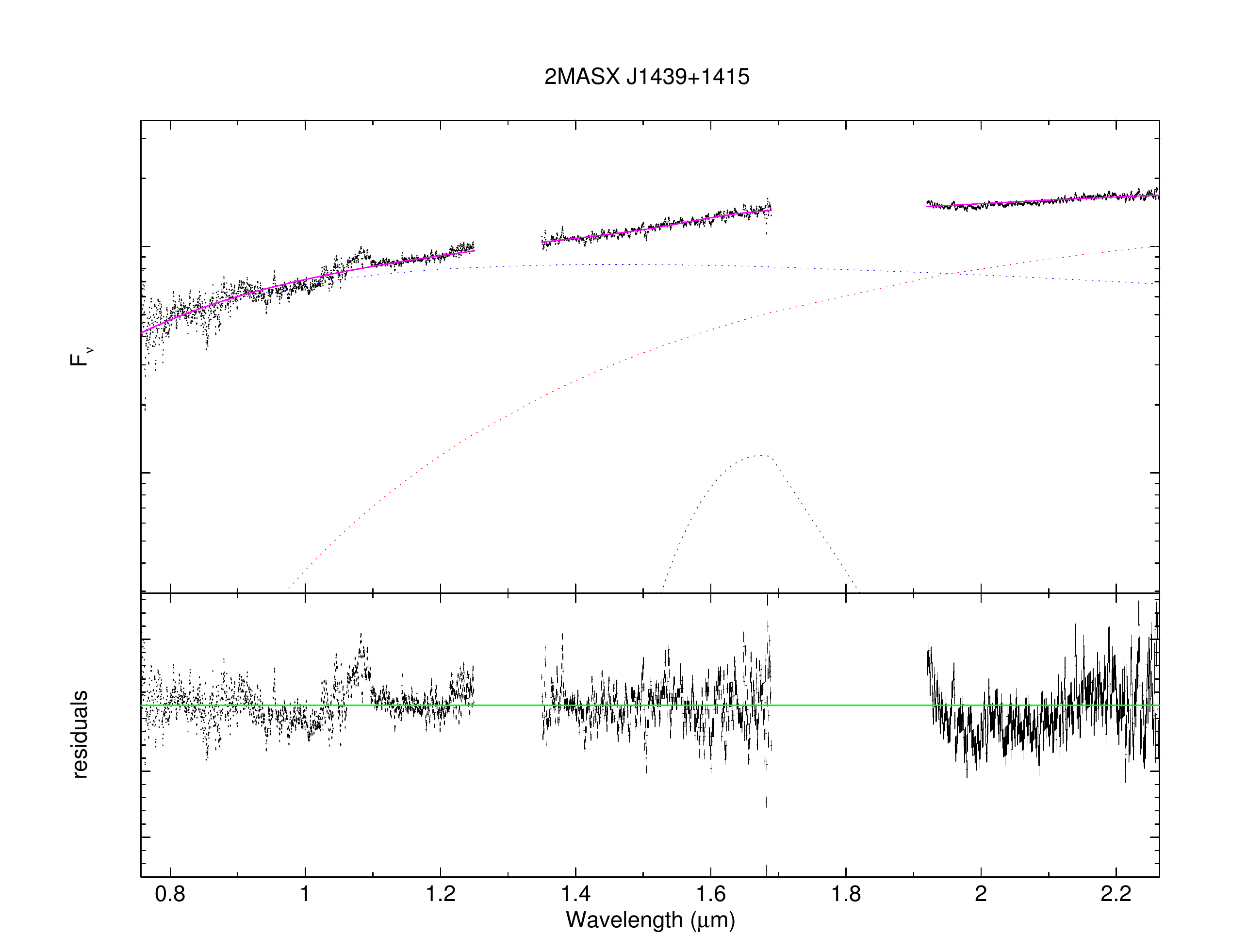}}
    \subfigure{\includegraphics[width=0.45\textwidth]{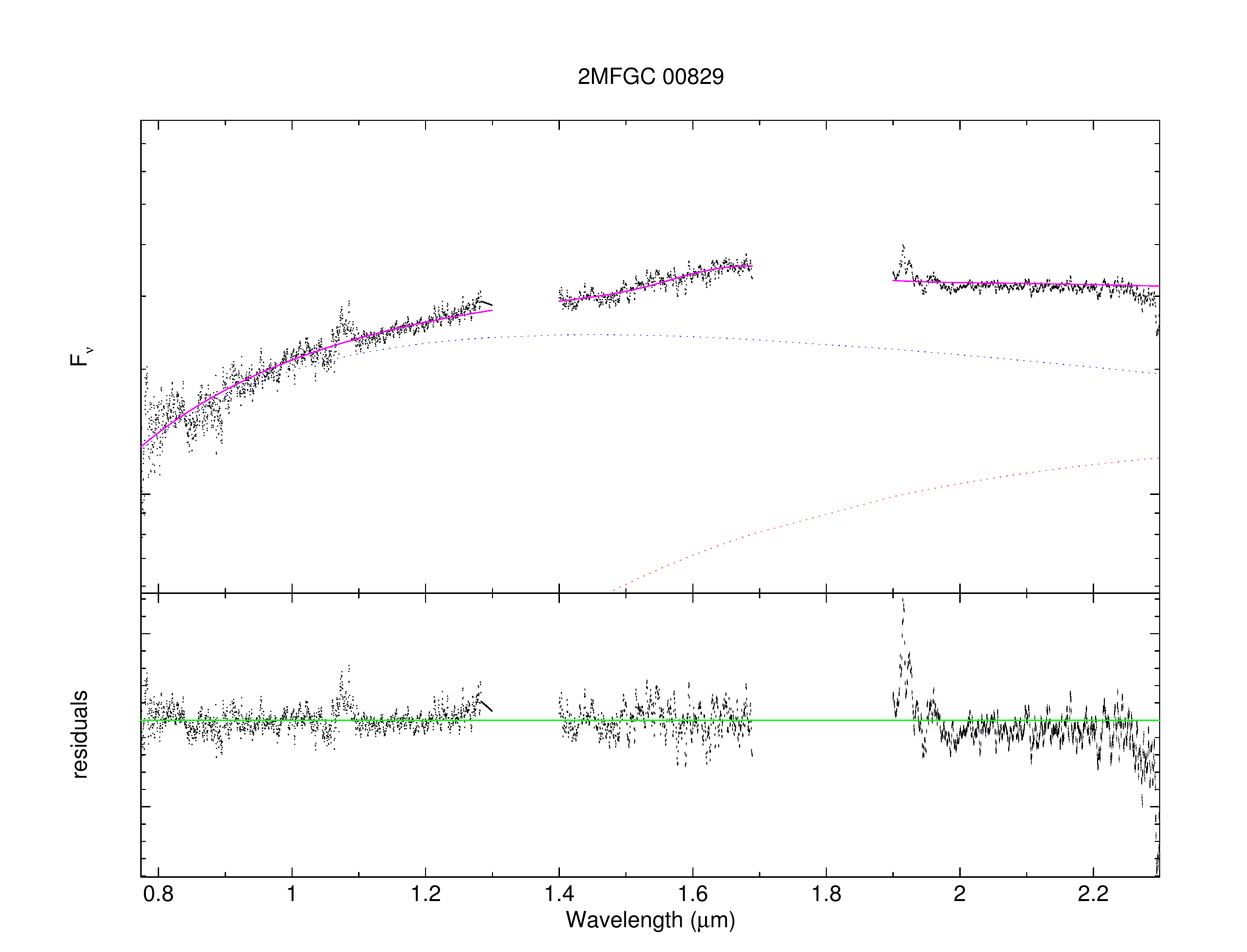}}
    \vfill
    \subfigure{\includegraphics[width=0.45\textwidth]{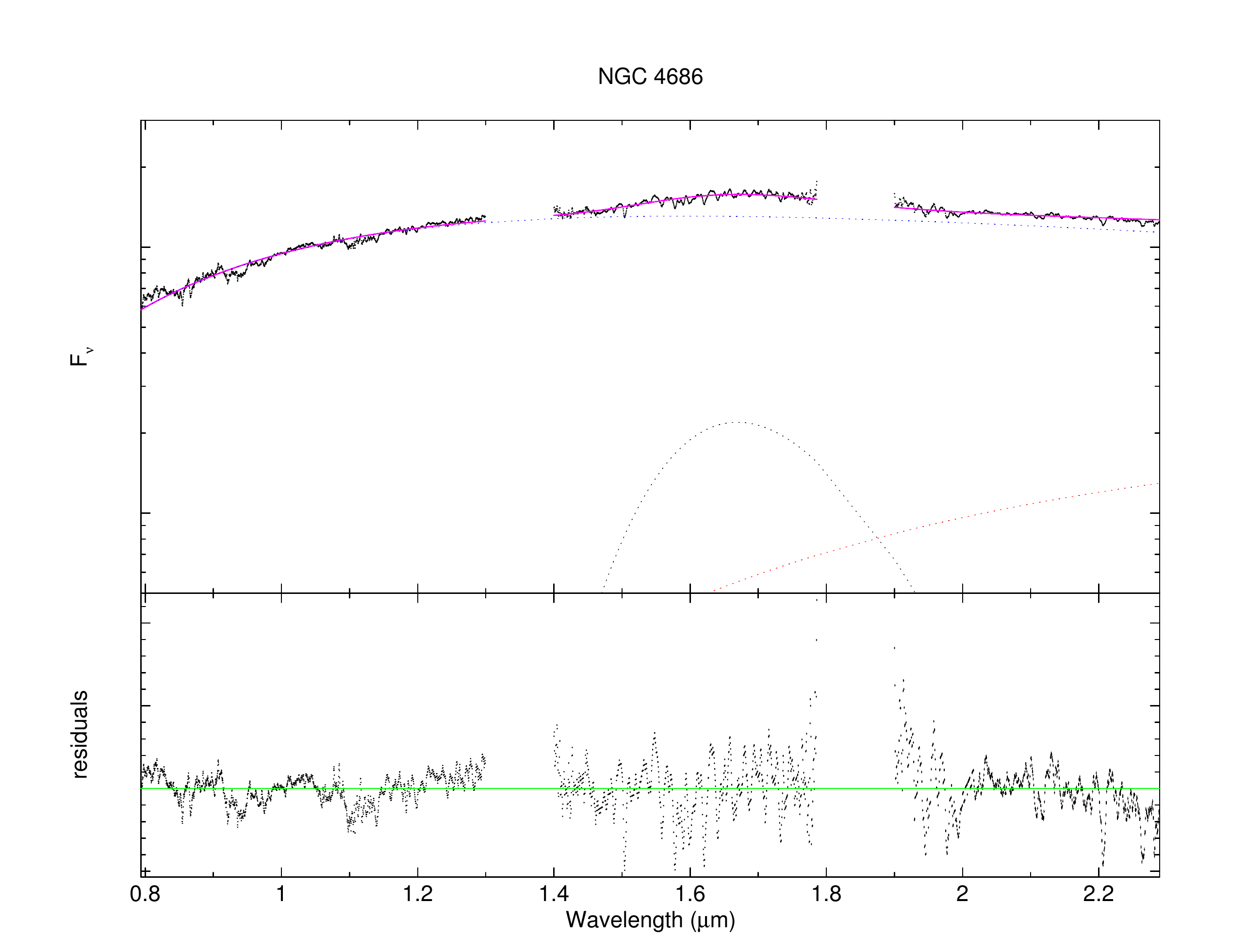}}
    \subfigure{\includegraphics[width=0.45\textwidth]{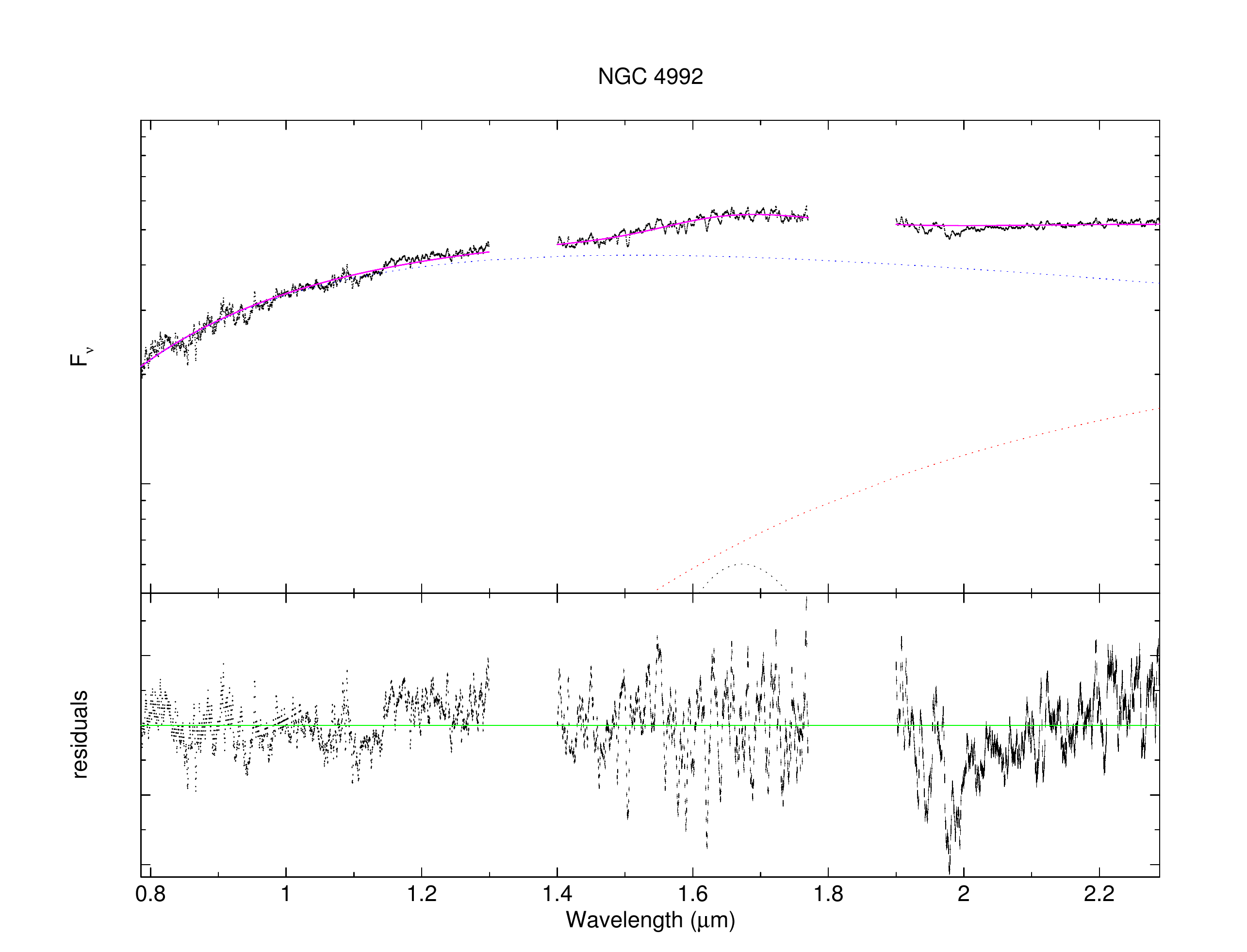}}
    \caption{Near infrared continuum fits and residuals of the four XBONGs. The fit consists of three components: a blackbody approximating the galaxy bulge's stellar spectrum (blue), a blackbody within the acceptable range of dust temperatures of $1300-1700$ K (red), and a gaussian component to account for the broad emission feature at 1.68~$\mu$m (shown in black if the contribution is large enough to appear on the plot). The fit is shown in magenta. }
    \label{fig:xbongfit}
\end{figure}

\begin{figure}
    \centering
    \subfigure{\includegraphics[width=0.45\textwidth]{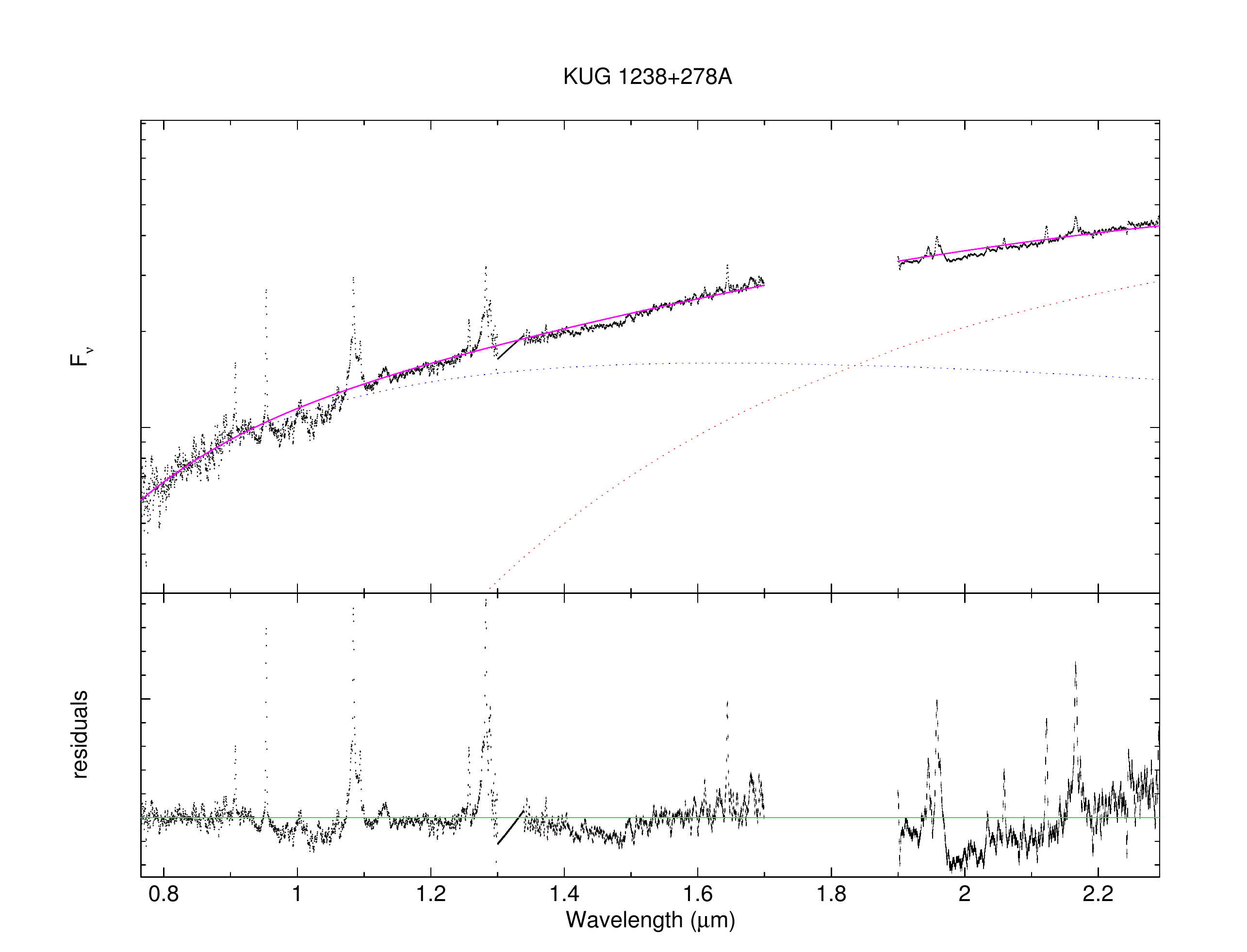}}
    \subfigure{\includegraphics[width=0.45\textwidth]{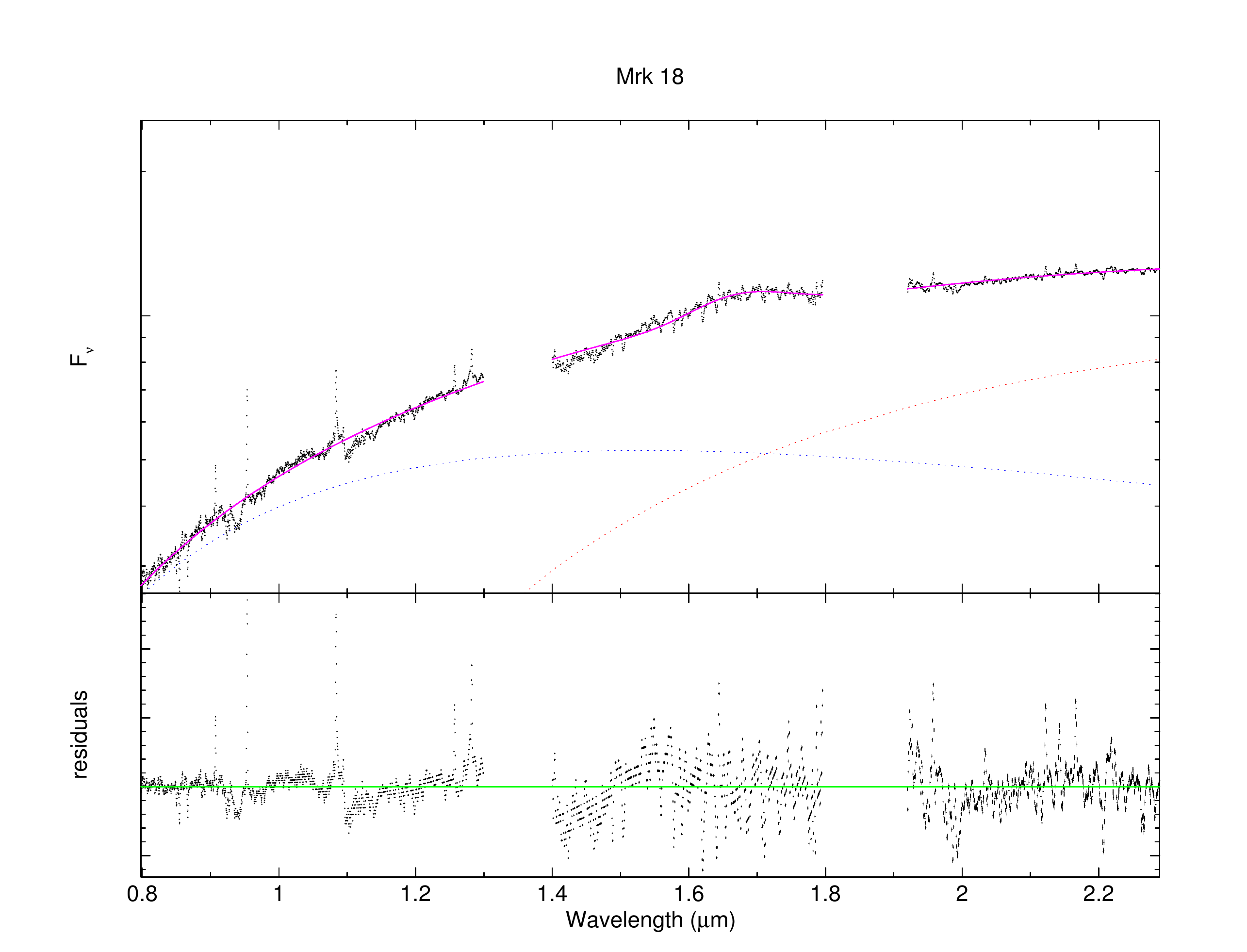}}
    \vfill
    \subfigure{\includegraphics[width=0.45\textwidth]{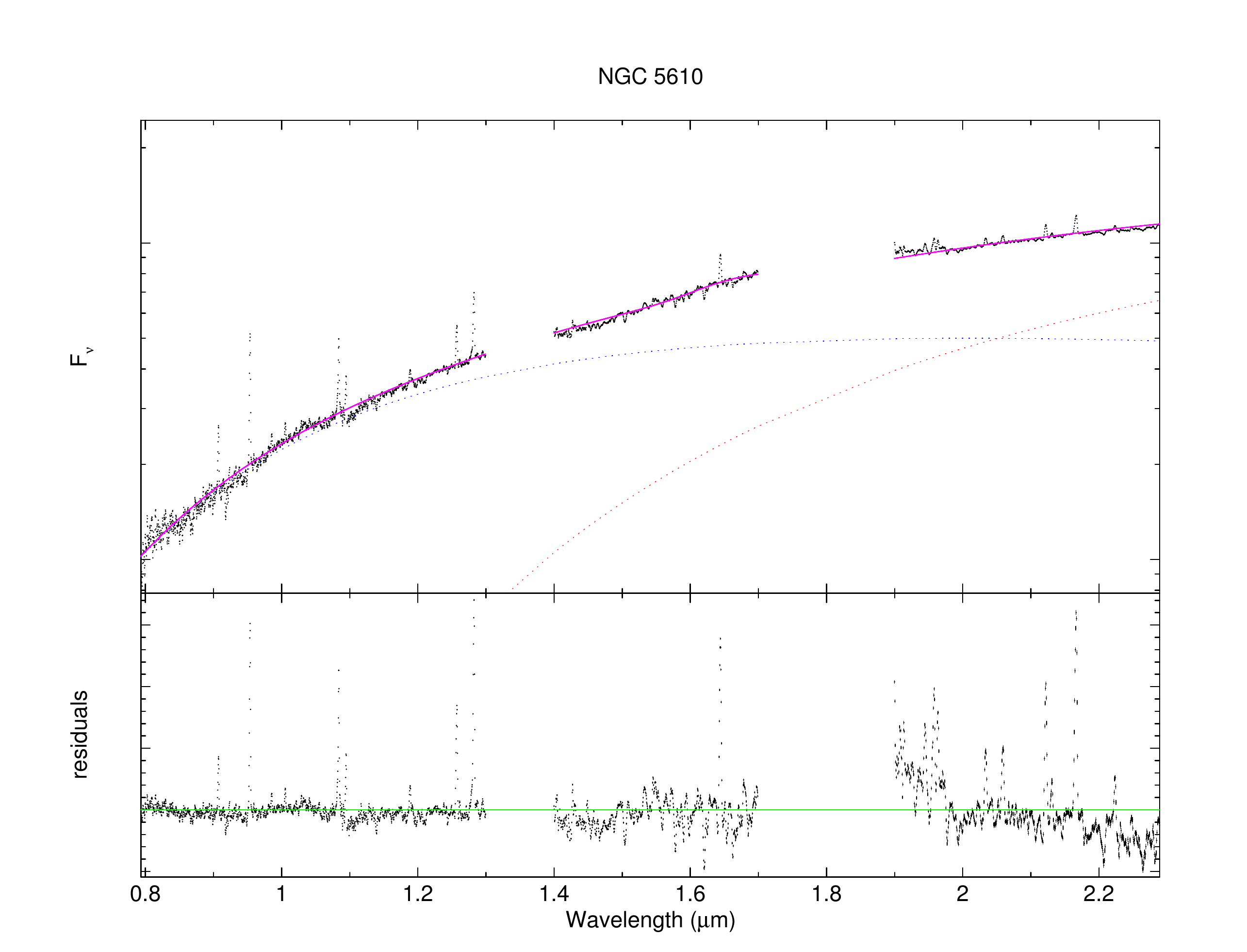}}
    \subfigure{\includegraphics[width=0.45\textwidth]{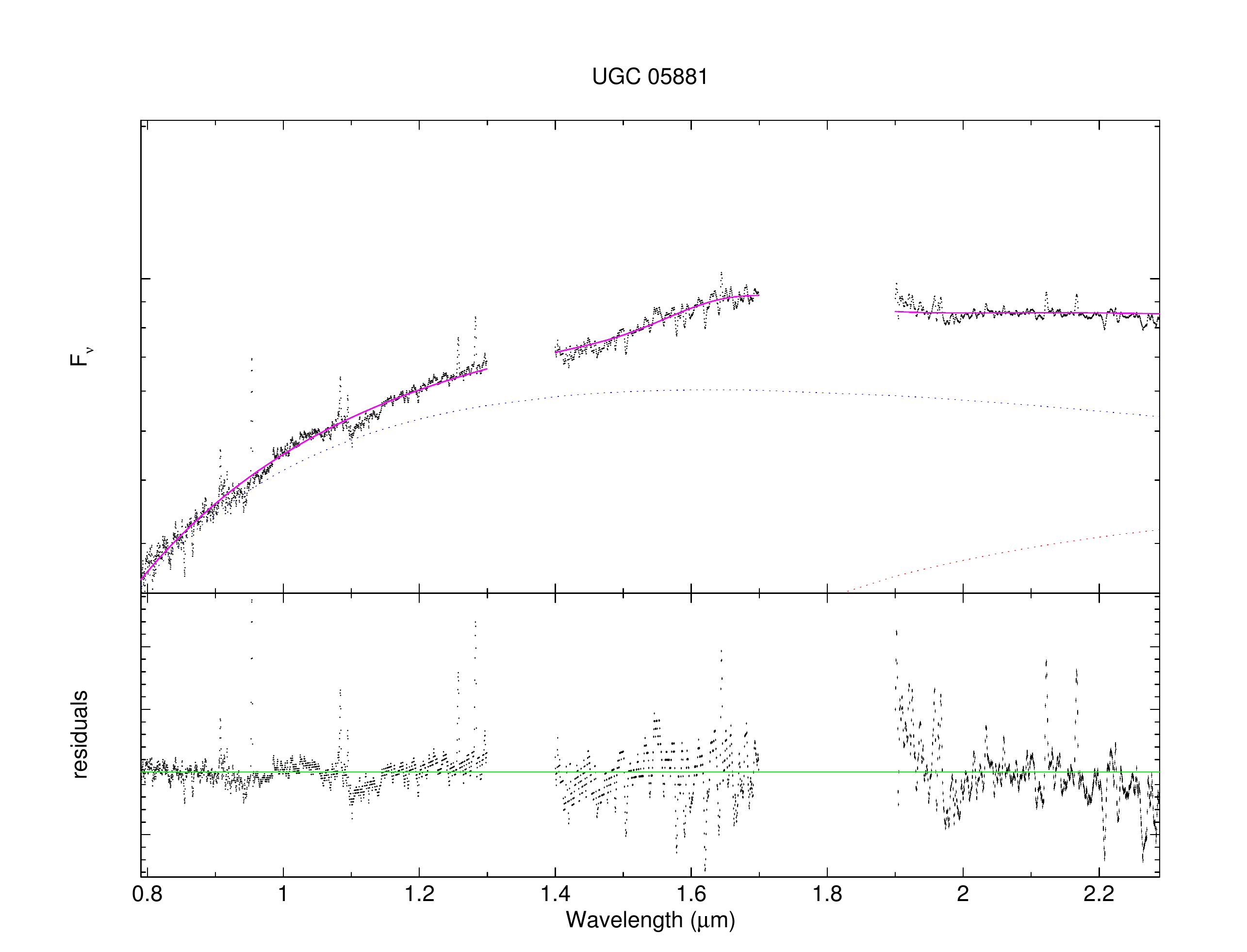}}
    \caption{Near infrared continuum fits and residuals of the four optically elusive AGN. The fit consists of three components: a blackbody approximating the galaxy bulge's stellar spectrum (blue), a blackbody within the acceptable range of dust temperatures of $1300-1700$ K (red), and a gaussian component to account for the broad emission feature at 1.68~$\mu$m (shown in black if the contribution is large enough to appear on the plot). The fit is shown in magenta. }
    \label{fig:oefits}
\end{figure}

Since our objects are not at high redshift (0.011 $< z <$ 0.071), the spectroscopic fiber does not contain a large fraction of the host galaxy. The average luminosity distance of our objects is $145\pm35$ Mpc. At this distance, the 3\arcsec~fiber of SDSS subtends $1.92\pm0.43$ kpc.  In Figure \ref{fig:images}, the red boxes indicate the regions over which the SDSS spectra were integrated (note that NGC 4686 is not a spectroscopic target in SDSS and 2MFGC 00829 is not in SDSS at all). All of the spectra were clearly taken from a nuclear region.

To quantify how much of the galaxy's starlight is included within the SDSS fiber, we have measured the ratio of the surface-brightness profiles of each galaxy integrated out to 1.5\arcsec~(half the width of the SDSS fiber) and a large radius of 50\arcsec. The percentage of the total galaxy light that is contained within the central 1.5\arcsec~radius for each galaxy is given in Table \ref{t:tab1}. Values range from $L_{1.5} / L_{tot} = 0.2-0.3$, except for NGC 4686, probably due to its being nearly edge-on, and 2MASX J1439+1415, since it is further away than most of our sources at $z=0.0708$. Note that this is the amount of the total galaxy light contained within the red squares in Figure \ref{fig:images}. Note also that for 2MFGC 00289 and NGC 4686, this information is not directly applicable because their optical spectra were not obtained from the SDSS. 

\subsubsection{$L_{opt} / L_{X}$ Distribution}
\label{sec:ox}

The starlight dilution paradigm was addressed in detail by \citet{moran02}. These authors simulated observations of distant \emph{Chandra} sources by integrating the host galaxy spectra of nearby, well-studied Type 2 AGN and deliberately diluting the spectra with the integrated light. They find that 60\% of their sources would not be classified as AGN at the average redshift of the deep \emph{Chandra} surveys, suggesting that starlight dilution is an important factor in the optical dullness of distant AGN.

\begin{figure}[ht]
\begin{center}
\plotone{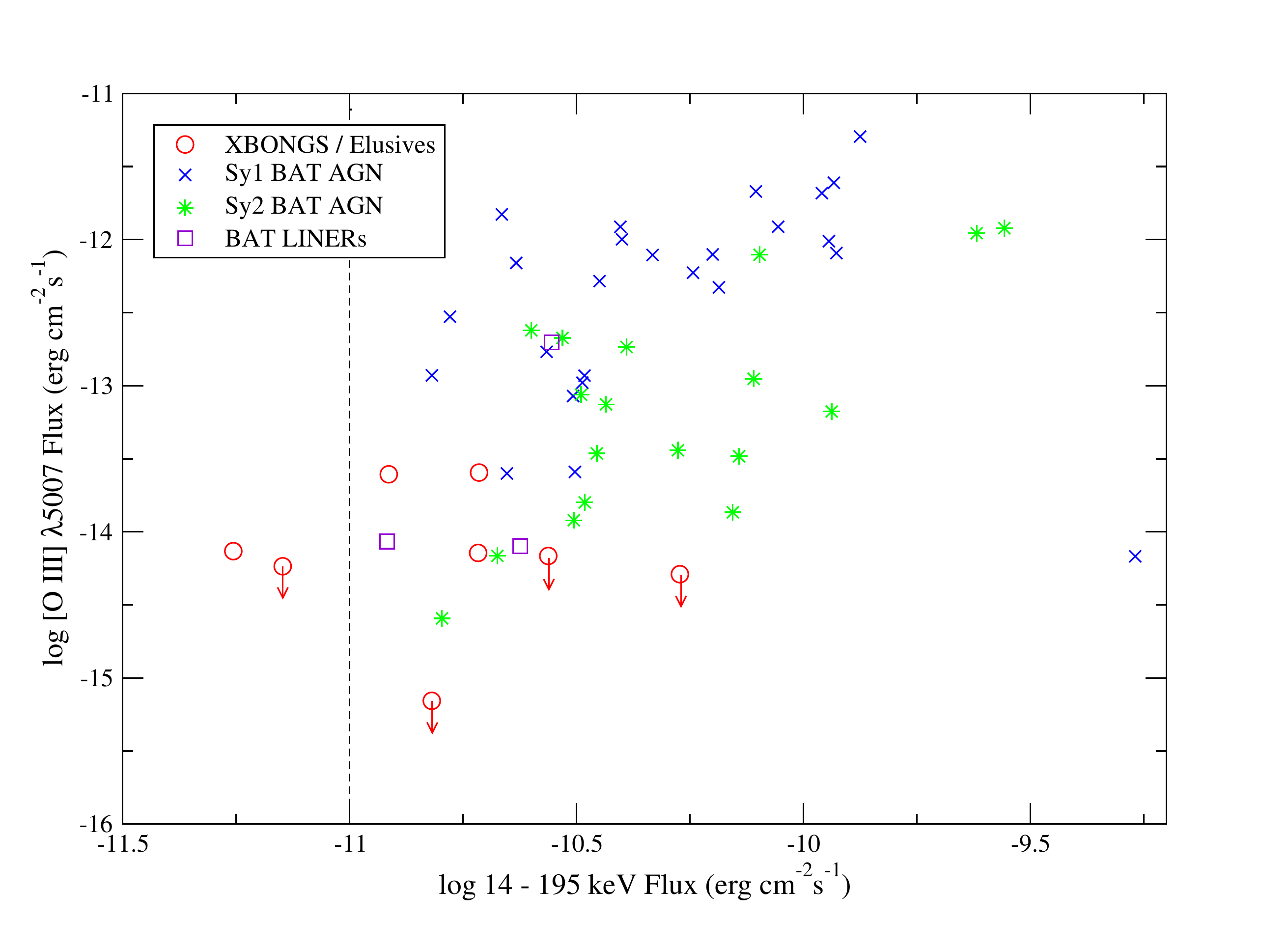}
\figcaption[]{The flux of the \oiii $\lambda$5007 emission line versus ultra-hard X-ray flux for subsets of the parent sample of BAT AGN and our sample of XBONGs and optically elusive AGN. The \oiii~flux values for the XBONGs are given as upper limits, since the line is not detected in these spectra. The vertical dashed line is the average flux limit of the \emph{Swift}-BAT survey. This figure may be compared with Figure 8 from \citet{cocchia07}. As in their sample, our XBONGs occupy a separate parameter space, and are not simply the tail end of the $L_{opt} / L_{X}$ distribution.
\label{fig:ratio}}
\end{center}
\end{figure}

Fortunately, \citet{cocchia07} have recommended a measure of optical dullness that should not be affected by galaxy starlight dilution, since it relies on direct measurements of line fluxes and not on equivalent widths. A key question about the nature of XBONGs is whether or not they are normal AGN that simply occupy the tail-end of the standard $L_{opt} / L_{X}$ distribution due to an abundance of starlight. In Figure \ref{fig:ratio}, we have plotted the flux in the \oiii~$\lambda~5007$ emission line versus the flux in the ultra-hard X-ray for our eight objects as well as for the overall \emph{Swift}-BAT AGN sample. Like the \citet{cocchia07} XBONGs, ours occupy a similar X-ray flux distribution as the majority of the sample, but have substantially different \oiii~fluxes. \citet{cocchia07} argue that this suggests the XBONGs are a ``truly distinct class" of AGN in terms of the ratio of optical and X-ray flux, and not just the tail end of the standard $L_{opt} / L_{X}$ distribution.

This conclusion, combined with the fact that the SDSS fiber does not contain a significant fraction of the exterior regions of the host galaxies, implies that starlight dilution, while certainly relevant to higher redshift samples as suggested by \citet{moran02}, is unlikely to be the cause of the optical normalcy of most of the galaxies in our sample.

The sole exception in our sample is 2MASX~J1439+1415, which we consider a viable case for starlight dilution. We note a strong resemblance to the starlight-diluted sample of \citet{moran02}; in particular the presence of H$\alpha$ but lack of H$\beta$ and very weak \oiii. Its greater distance ($z=0.0708$) places a large fraction of the stellar light within the SDSS spectroscopic fiber. 

\subsubsection{4000 \AA~Break}
\label{sec:4000a}

The 4000~\AA~break is a steep drop off of radiation blueward of 4000~\AA, the combined effect of a dearth of hot blue stars and blanket absorption of high energy radiation by metals. It is a ubiquitous feature in galaxy spectra. The strength of the break feature is a tracer of the contribution to the spectrum of galaxy starlight versus nuclear emission, and is given by:

\begin{equation}
\Delta_{4000} = \frac{F^{+}-F^{-}}{F^{+}}
\end{equation}

where $F^{+}$ is the average flux density between 4050~\AA~and 4250~\AA, and $F^{-}$ is the average flux density between 3750~\AA~and 3950~\AA, as put forth by \citet{bruzual83}. Both are measured in the object's rest frame. The values of $\Delta_{4000}$ for each object are given in Table~\ref{t:tab1}.

If nuclear emission is dominant, the value of $\Delta_{4000}$ is very low or negative. The higher the value of $\Delta_{4000}$, the greater the contribution of host galaxy starlight. 

As stated in the introduction, \citet{cacc07} conclude that starlight dilution is the primary driver behind the optical dullness of their sample of elusive AGN from the \emph{XMM-Newton} bright serendipitous survey. They show that the strength of the break is anticorrelated with the unabsorbed X-ray luminosity. We can measure the 4000~\AA~break strength for all of our objects except NGC 4686, since its optical spectrum does not extend blueward far enough. 

\begin{figure}[ht]
\begin{center}
\plotone{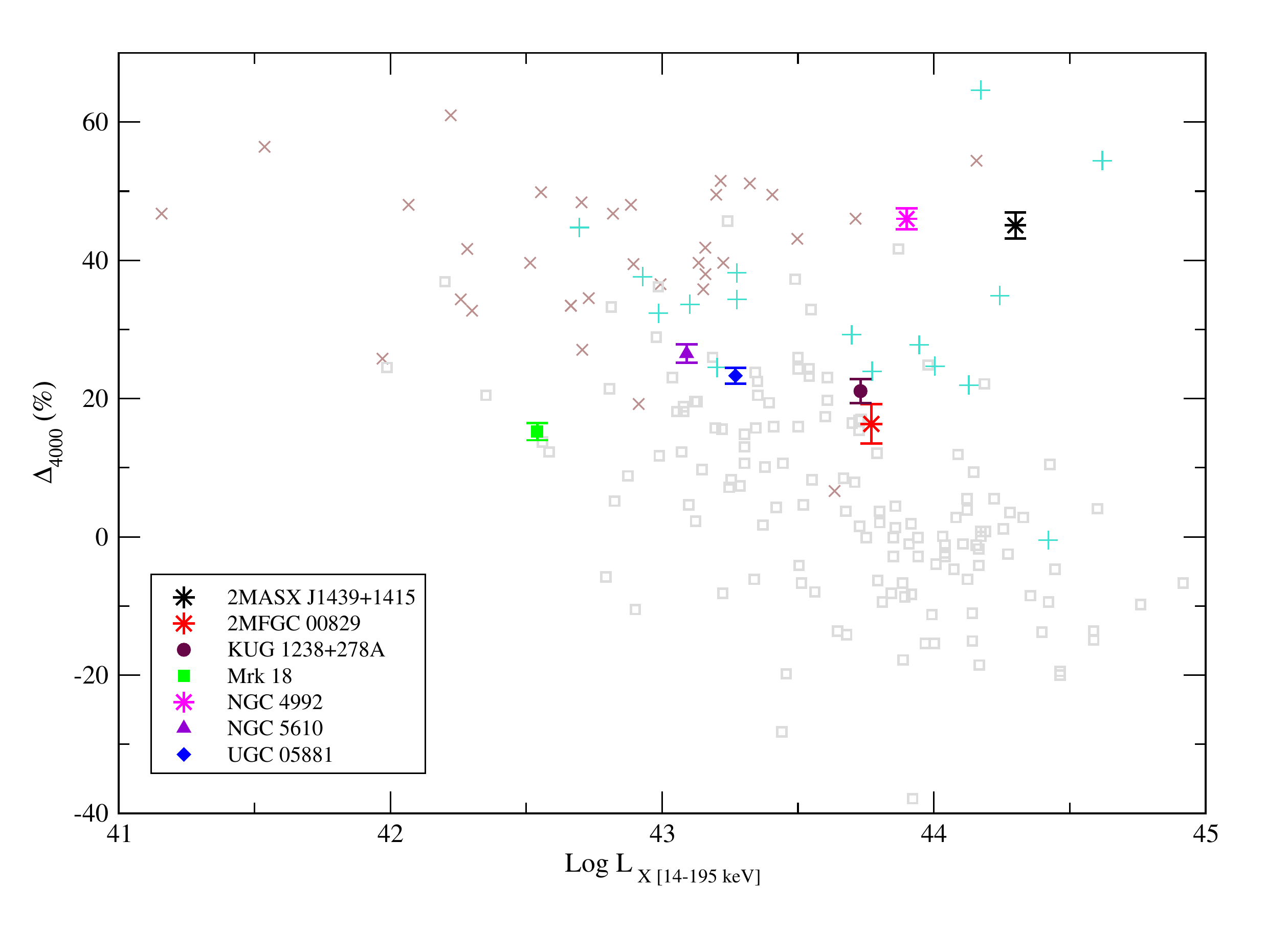}
\figcaption[]{Position of the XBONGs (shown as stars) and optically elusive AGN (filled symbols; same as in previous figures) on the anticorrelation between the strength of the 4000 \AA~break ($\Delta$) and the unabsorbed X-ray luminosity reported by \citet{cacc07}. Their sample includes Type 1 AGN (grey squares), Type 2 AGN (turquoise plusses) and elusive AGN (brown crosses). Two of our objects have an indication of heavy starlight dilution (i.e., $\Delta_{4000} \gtrsim 40$\%).
\label{fig:4000a}}
\end{center}
\end{figure}

In Figure \ref{fig:4000a}, we reproduce Figure 4 of \citet{cacc07}, including our objects. To scale their de-absorbed X-ray luminosities from the \emph{XMM-Newton} $2-10$~keV range to our range, we employ a rough scaling factor of $\sim2$, obtained by integrating a power law slope of approximately 1.7~photons~cm$^{-2}$~s$^{-1}$~keV$^{-1}$ between the two bands. Immediately obvious is that our sample is not concurrent with their sample of elusive AGN. This is most likely because their sample has an average redshift of $z=0.184$, considerably higher than ours, allowing more of the host galaxy light to occupy the slit. Our objects occupy a similar X-ray space as their optically elusive AGN, but are lower in $\Delta_{4000}$ space and thus have less galaxy contribution to the optical spectrum. 

Two objects are outliers: the XBONGs NGC 4992 and 2MASX J1439+1415. In the case of NGC 4992, we believe the galaxy contributes strongly due to extreme dust obscuration near the nucleus. The position of 2MASX J1439+1415 supports our theory that this object is optically normal because of starlight dilution, since it has a high value of $\Delta_{4000}$ and a high X-ray luminosity, similar to the starlight-diluted Type 2 AGN in the diagram.

\subsection{Star Formation Activity}
\label{sec:sf}

The optically-elusive AGN, which exhibit emission line ratios characteristic of star-forming HII regions, could be explained by an excess of star formation in the host galaxy, or a circumnuclear starburst. If the nuclear emission is relatively weak, such a starburst could outshine it, and cause the line ratios to indicate HII region activity rather than activity from the central engine. 

We note here that one should use caution when considering star formation and the infrared line ratios presented in Section \ref{sec:linerat}. The emission used in that diagnostic could potentially be excited by supernovae or UV radiation from O and B stars (see Section \ref{sec:linerat}). 

To determine the degree to which star formation is affecting our sample, we obtained $U-R$ photometry from SDSS for the seven of our objects in that database (this excludes 2MFGC 00829). The numbers are given in Table \ref{t:tab1}. The average for our sample is $U-R=2.42\pm0.15$. We assembled a control sample of Type 2 Seyferts selected by the optical line ratio diagnostics from \citet{kewley06}, and selected smaller control samples for each object based on total stellar mass. For this control sample of matched stellar mass, the average is $U-R=2.41\pm0.12$, the same as for our sample. We conclude that our objects are not systematically bluer than a control sample of similar $M_*$, and are thus unlikely to be seriously affected by star formation. However, if our objects are highly dust obscured, there could still be significant star formation detectable only in the re-radiated infrared. Additionally, SDSS photometric magnitudes are taken for the integrated galaxy within the Petrosian radius, not only for the nucleus. Since our spectra are mostly nuclear, it is of interest to see whether the photometry of the central regions might indicate a circumnuclear starburst.

The SDSS also makes available the ``fiber magnitudes" for each photometric object, which are the magnitudes of the regions of the galaxy inside the 3\arcsec~SDSS spectroscopic fiber (i.e., the magnitudes from the region inside the red boxes in Figure \ref{fig:images}). These values are more sensitive to circumnuclear starbursts than whole-galaxy photometry. For the fiber magnitudes, we find for our sample that $(U-R)_{fiber} = 2.49 \pm 0.17$. For the same control sample as above, we find $(U-R)_{fiber} = 2.44 \pm 0.02$. Even for the fiber magnitudes, there is no difference between our galaxies and a matched-$M_*$ control sample of Type 2 AGN. Interestingly, both our sample and the control sample were, on average, redder in the fiber magnitudes than for the integrated photometry. 

The detection of prominent PAH features in the MIR \emph{Spitzer} spectra would also indicate prominent star formation. Of our three \emph{Spitzer} objects, only one, Mrk 18, exhibits PAH emission; in particular, the PAH lines at 6.2, 7.62, 8.6, 11.3 and 12.7$\mu$m are present. \citet{draine84} postulated that these are likely caused by bending and stretching of complex carbon molecules in star forming regions. The strong detection of these lines in Mrk 18 imply a nuclear starburst in this object, which could very well be why optical line ratios classify it as an HII region instead of an AGN. 

An independent measurement of the star formation rate (SFR) can be obtained for those of our objects with \emph{Galaxy Evolution Explorer} (GALEX) far-UV data. \citet{salim07} calculated an empirical conversion factor between the SFR and the FUV flux (see their Equation 10).  Using this factor, we can calculate lower limits of the SFRs in our objects. To get true values, we would need to apply an extinction correction. At present, such extinction corrections to the FUV flux are calibrated only for normal, non-AGN galaxies, and such an analysis for hybrid star-forming galaxies and AGN exceeds the scope of this work. Additionally, although it is unlikely that a Type~2~AGN would have any continuum emission in the UV, it is not impossible that there is some contribution to the \emph{GALEX} FUV flux from the AGN itself. These lower limits on the SFR are given in Table~\ref{t:tab1}. The typical value is $\sim0.1$~M$_\odot$~yr$^{-1}$, which is consistent with the results in \citet{salim07} for Type~2~AGN. Note that these SFRs were calculated from \emph{GALEX} photometry for the entire galaxy, not exclusively the nuclear regions from which the optical spectra were taken.

\subsection{Radiatively-Inefficient Accretion Flows}
\label{sec:riaf}

Below a certain accretion rate, an optically-thick accretion disk at small radii fails to form, and gives way instead to a geometrically thin disk at large radii, with a very hot radiatively inefficient accretion flow taking over near the black hole. \citet{yuan04} postulate that XBONGs are a consequence of such a configuration. Since the classical accretion disk is responsible for the widespread ionizing radiation that produces the AGN emission lines, it is possible that an object with a radiatively inefficient accretion flow (RIAF) might not exhibit emission lines. We then must check that our sample does not have unusually low accretion rates. We obtained stellar velocity dispersions ($\sigma_*$) from the line widths in the optical SDSS spectra to calculate black hole mass, following the method of \citet{gultekin09}:

\begin{equation}
log~\frac{M_{BH}}{M_\odot}= (8.12\pm0.08)+(4.24\pm0.41)\times~log~(\frac{\sigma_*}{200~\rm{km~s}^{-1}}).
\end{equation}

 From these masses, we calculated the Eddington luminosity, $L_{Edd} = 1.38\times10^{38}(\frac{M}{M_\odot})\rm~erg~s^{-1}$. Two objects are not included in SDSS: NGC~4686, for which we do have an optical spectrum from Kitt Peak Observatory, and 2MFGC~00829, which is in the 6DF survey and for which we have a spectrum from the SNIFS instrument on the University of Hawaii 88 inch telescope. For 2MFGC~00829, we obtain the width of the \oiii~$\lambda~5007$ emission line using the IRAF routine SPLOT. We then use the relationship between the $FWHM_{[O III]}$ and the stellar velocity dispersion given by \citet{gaskell09} to obtain $\sigma_*$, and again calculate the black hole mass using \citet{gultekin09}. NGC~4686 has no emission lines in the optical spectrum, so we cannot obtain a black hole mass or Eddington ratio for this object. As a surrogate for bolometric luminosity, we use the hard X-ray luminosity multiplied by a correction factor of 15, as suggested by \citet{vasudevan10}. Our calculated Eddington ratios are shown in Table \ref{t:tab1}. 

The calculated Eddington ratios for the objects is on average 0.05,  excluding the one clear outlier 2MASX~J1439+1415 with the high value of $L_{bol}/L_{Edd} = 0.44$. The average value of the sample is not discrepant with normal Type~2~AGN according to \citet{trump11}, and is above their threshold for RIAF accretion of $L_{bol}/L_{Edd} \ll 10^{-2}$. From this, we conclude that the explanation for the absence of optical lines in our sample is not a radiatively-inefficient or advection-dominated accretion flow.

\section{Conclusion}
\label{sec:conclusion}

We have presented near-infrared and optical spectra and archival MIR spectra of four optically elusive AGN and four XBONGs. We have shown that the majority of AGN selection techniques have varied and limited success at selecting these objects as AGN, despite the fact that extinction in the infrared is much less important than in the optical. The XBONGs continue to have no emission lines in the NIR, while the optically elusive AGN do have NIR emission lines; additionally, the line ratios in the NIR fall in the AGN region of diagnostic plots, indicating that we can detect nuclear excitation in the NIR spectra. Unfortunately, these NIR line ratio diagnostics do not cleanly distinguish between HII region and AGN samples, and so are not  useful for unambiguously selecting AGN.

 We do not detect coronal emission lines in the NIR in any of our objects, but all three of our objects with archival MIR \emph{Spitzer} spectra do exhibit at least one coronal emission line. One optically elusive AGN, KUG~1238+278A, has a hidden broad-line region exposed in the NIR. Infrared photometric AGN selection techniques recover three of our objects:  the hidden BLR object KUG~1238+278A, the starlight dilution candidate 2MASX~J1439+1415, and, surprisingly, the heavily obscured NGC~4992.

We have shown that the lack of optical indicators of AGN activity in our sample is unlikely to be explained by a high star formation rate overwhelming the nuclear emission, since our sample does not have $U-R$ colors significantly different than a control sample with similar $M_*$ in either whole-galaxy or nuclear photometry. We have also shown that our objects do not have especially low accretion rates, and so are not likely to have truncated accretion disks with radiatively-inefficient accretion flows. Starlight dilution in the SDSS fiber is not likely for most our objects, since they are all very nearby. Additionally, our objects do not occupy the tail-end of the $L_{opt} / L_{X}$ distribution, as one would expect for a starlight-diluted sample, and analysis of their 4000~\AA~break strength indicates that most objects' spectra do not have a large galaxy contribution. The exception is 2MASX~J1439+1415.

We find that the most likely explanation for the majority of the objects in our sample is dust obscuration, as in the case of NGC 4992. The sample on average has an $E(B-V)$ reddening an order of magnitude higher than a control sample. The XBONGs for which we have MIR \emph{Spitzer} information exhibit silicate absorption features indicative of a high dusty column density. We have compared the morphologies of our objects to the other \emph{Swift}-BAT AGN and have found tentative indications of a higher population of bars, which could channel dust into the central regions, but refrain from conclusions due to our small sample size.

In short, no one infrared AGN selection test succeeds in recovering all or even most of the objects. Our small sample of hard X-ray selected AGN contains objects that are optically normal for a variety of individual reasons, including starlight dilution, nuclear starbursts and dust obscuration. We do not find a blanket cause for optical normalcy. As the BAT sample size continues to increase and the survey becomes more sensitive, more XBONGs selected in the ultra-hard X-ray will be discovered that can be subjected to these tests. This will lead to a better understanding of what AGN are missed by optical and NIR selection techniques. Finding ways to address the large incompleteness of optical samples is vital, since these samples are frequently used in cosmological and galaxy evolution studies.

\acknowledgments

M.K. acknowledges support from Swiss National Science Foundation (NSF) grant PP00P2 138979/1.
We thank Keith Arnaud of NASA Goddard Space Flight Center for his help in adapting XSPEC for use on NIR spectra. We would also like to thank an anonymous referee for suggestions that improved the paper.

The authors wish to recognize and acknowledge the very significant cultural role and reverence that the summit of Mauna Kea has always had within the indigenous Hawaiian community.  We are most fortunate to have the opportunity to conduct observations from this mountain.
 
\appendix
\section{Appendix: Discussion of Individual Objects}
\label{sec:onebyone}
\subsection{2MASX J1439+1415}
\label{sec:2mas}

This is our highest redshift object: the SDSS spectroscopic fiber contains 62\% of the total galaxy light. It has a high Eddington ratio, low X-ray column density, and an elliptical morphology which generally implies a small dust component. The optical spectrum shows a striking resemblance to the starlight-diluted artificial spectra of \citet{moran02}; in particular, the presence of H$\alpha$ but lack of H$\beta$ and very weak \oiii. While these features are also present in the other XBONG spectra, those objects have high X-ray column densities and low Eddington ratios, and are much nearer, so the spectroscopic fiber encompasses only $\sim$20\% of the total galaxy light. It passes all three NIR photometric tests. Additionally, its 4000~\AA~break strength is quite high, consistent with heavy starlight dilution. Finally, its NIR continuum fit resembles the optically elusive AGN in our sample, rather than that of its fellow XBONGs. We conclude that this object is most likely optically normal due to dilution of the spectrum by host galaxy starlight; the object would likely be classified as a Type 1 AGN if it were nearer or the spectrum excluded more of the host galaxy.

\subsection{2MFGC 00829}

This XBONG is difficult to study because of a lack of information; it is not in the SDSS, and so we do not have its $U-R$ color. The only available image is from 2MASS, which is lower resolution. The low quality X-ray spectrum prevents us from constraining its line-of-sight column density. There is no \emph{Spitzer} data for this object, so we cannot look for silicate absorption troughs. Since it is not at high redshift, we think starlight dilution is unlikely. Since it is an XBONG, not an elusive AGN, and has no emission lines at all, nuclear star formation is not an explanation. Our best guess for this object is dust obscuration, perhaps due to widespread dust in the host galaxy, since it appears to be an edge-on spiral. The fit of its NIR continuum is consistent with the dust-obscured XBONGs, in that the hot nuclear dust component is quite small compared to the stellar blackbody component, implying further obscuration of the nucleus but a cold dust component further away. 

\subsection{KUG 1238+278A}

This object's NIR spectrum is dominated by nuclear emission. We see broad Paschen lines and the flat continuum associated with AGN, a classical hidden broad line object. The reason that it falls in the star forming region of optical diagnostic plots may be due to a nuclear starburst, possibly fueled by an ongoing merger; its image suggests it may be morphologically disturbed by an interaction. It is classified as a compact galaxy in the 2MASS NIR galaxy morphology atlas of \citet{jarrett00}, which they describe as ``not obviously spiral or elliptical." Its NIR continuum fit is similar to the other optically elusive AGN, consisting of relatively equal contributions from a stellar blackbody and a hot dust component. It passes all three NIR photometric tests. Dust obscuration must play some role, because if we penetrate the dust by observing in the NIR, its AGN nature becomes clear; however, this dust obscuration must be weaker than that affecting the XBONGs, since their NIR spectra remain those of inactive galaxies. 

Interestingly, we note that this object has a candidate ``voorwerp" nearby, from the Galaxy Zoo survey for AGN-ionized gas clouds by \citet{keel12}. Those authors state that it would be quite interesting to see the gas cloud being ionized by an AGN when no AGN was clearly present. KUG 1238+278A is such a case; in the optical, the only indication of AGN activity in this object is its voorwerp. 

\subsection{Mrk 18}

Mrk 18 is a previously unknown AGN, first discovered in the hard X-ray, and \citet{weaver10} suggest this is because of ``extinction of the optical emission lines." \citet{winter10} classify it as an HII region / LINER, as we do. Our strongest candidate for nuclear starburst activity, this object exhibits strong PAH features associated with star formation in its \emph{Spitzer} spectrum. It is unlikely to be seriously affected by dust obscuration, since its $E(B-V)$ value is quite low (0.05) and the \emph{Spitzer} spectrum does not show silicate absorption troughs. It has numerous AGN diagnostics in the infrared, including the detection of MIR high ionization lines \nev~and \oiv, as well as NIR emission line ratio diagnostics. Finally, Mrk~18 falls in the hybrid SB/AGN portion of the \citet{genzel98} MIR classification diagram. We believe that, rather than heavy extinction, this object's line ratios are affected by a nuclear starburst, causing it to fail classical optical AGN ratio tests. Some nuclear dust may slightly reduce the AGN contribution to the optical lines with respect to the NIR lines, since the object passes NIR line diagnostics. 

\subsection{NGC 4686}

This XBONG is almost certainly optically normal due to dust obscuration. \citet{ajello09} classify this object as a radio-quiet AGN of the XBONG type. It has a high X-ray column density and its \emph{Spitzer} spectrum exhibits silicate absorption, although not to the same extreme degree as NGC~4992. We do not significantly detect \nev~in the NIR spectrum, but we do detect \oiv~in the MIR spectrum (see Figure \ref{fig:spitzer}), which is an unambiguous AGN indicator. Since NGC~4686 is not an SDSS spectroscopic target and its spectrum does not extend blueward enough to contain the 4000~\AA~break, we cannot estimate the amount of starlight dilution present in our optical spectrum from Kitt Peak. The NIR spectrum, like the optical, has no emission lines and the NIR continuum fit is consistent with a stellar population with a minimal hot dust contribution, probably shielded by colder dust. The obscuration may be due to host galaxy dust, since a dust lane is clearly present in the image.

\subsection{NGC 4992}
\label{sec:4992}

\citet{asmus14} nicely summarizes the literature on this object; in short, it has multiple conflicting optical classifications ranging from totally inactive \citep{masetti06} to Type 1 AGN \citep{veron10}. \citet{saz05} show that the X-ray properties imply an obscured AGN. We do detect MIR \oiv~emission in this object, which is an AGN indicator, and it passes two NIR photometric tests for reasons that are unclear. The \emph{Spitzer} spectrum shows very prominent silicate troughs, which dominate the continuum. Such deep troughs are typical of AGN/starburst composite objects in the \citet{asmus14} MIR atlas of nearby AGN, but NGC 4992 lacks the strong PAH features characteristic of such composites. However, the PAH emitting regions in this object may be blocked by dust, with the rest of the nuclear regions. It has the highest X-ray column density of any of our objects. All indicators suggest that this is AGN is heavily dust obscured.

\subsection{NGC 5610}

This face-on barred spiral galaxy is optically elusive and classifies as a starburst / HII region on optical line ratio diagnostic diagrams. Like with NGC 4686, \citet{ajello09} call it a radio-quiet AGN of the XBONG type. We do not believe starlight dilution is a viable explanation for the optical normalcy of this object for several reasons: it is nearby ($z=0.0168$), the SDSS spectrum is quite nuclear, and the strength of the 4000~\AA~break is consistent with Type 2 AGN and even some Type 1 AGN (see Figure \ref{fig:4000a}). It does not have an unusually low Eddington ratio. Since we do not have a MIR spectrum for this object, we do not know if it exhibits PAH features or silicate absorption. Its face-on orientation, however, precludes dust absorption on large scales in the host galaxy along the line of sight. So, this object possibly hosts a nuclear starburst, the emission lines from which dominate the emission lines from gas ionized by the AGN itself due to dust blocking the nucleus. 

\subsection{UGC 05881}

In addition to the \emph{Swift}-BAT survey, UGC 05881 was also detected by the IBIS instrument on \emph{INTEGRAL}. \citet{masetti10} classify the \emph{INTEGRAL} source optically as a low-ionization nuclear emission line region (LINER) and as Compton-thin. Interestingly, those authors claim that there is no reddening local to the AGN host, and thus declare it a ``naked" LINER, which is the first of its kind. The optical classification of LINER is not entirely inconsistent with our optical classification as an HII region, since the object falls in the composite regime between star formation and AGN-dominated; i.e., between the \citet{kewley06} and \citet{kauffmann03} demarcations. We do, however, find that there is local reddening on the order of $E(B-V) \sim 0.25$. The object is successfully classified as an AGN using NIR emission line diagnostics; therefore we believe there is some slight nuclear dust obscuration causing it to fail optical AGN diagnostics. Interestingly, the NIR continuum fit for this object resembles the XBONGs, rather than the optically elusive AGN, in that it is best matched by a predominant stellar component and only a small hot dust contribution. Perhaps the slight obscuration of this object is due to large-scale colder dust in the central regions of this barred spiral galaxy. 

Note that \citet{masetti10} refers to this object by its other name, MCG~$+04-26-006$.

\clearpage


\tabcolsep=0.05cm
 \begin{deluxetable}{lccccccccccc}
 \label{tab:list}
 \tablewidth{0pt}
 \tabletypesize{\scriptsize}
 \tablecaption{XBONGs and Optically Elusive AGN\label{t:tab1}}
 \tablehead{
 \colhead{Object} &
 \colhead{Class} &
 \colhead{Redshift} &
 \colhead{Eddington} &
 \colhead{$Log(L_X)$} &
 \colhead{$Log (N_H)$} &
 \colhead{$E(B-V)$} &
 \colhead{$U-R$} &
 \colhead{SFR$_{min}$} &
 \colhead{ Light} &
 \colhead{Morphological} &
 \colhead{$\Delta_{4000}$} \\
 \colhead{ } &
 \colhead{ } &
 \colhead{ } &
 \colhead{Ratio } &
 \colhead{ erg/s } &
 \colhead{ cm$^{-2}$ } &
 \colhead{ } &
 \colhead{ } &
 \colhead{M$_\odot$~yr$^{-1}$ } &
 \colhead{in Fiber} &
 \colhead{Type} &
 \colhead{ }}
 
 \startdata 
 
2MASX J1439+1415	 	 &     XBONG	& 0.0708	&  0.441  & 44.28 & 22.32 &  -  & 2.9 & - & 0.62 & E? & 45.05 \\
2MFGC 00829      			 &	XBONG	& 0.0467 	&   0.010  &  43.45  & - &   - &  -  & 0.15 &  0.34 & S? & 16.33 \\
KUG 1238+278A			 &	Elusive	& 0.0566	&   0.049 & 43.74 & - & 0.266 & 1.96 & 0.74 & 0.25 & Pec/Int & 21.07 \\
Mrk 18					 &	Elusive	& 0.0111	&   0.024 & 42.53 & 22.93 & 0.050 & 1.86 & 0.12 & 0.21 & Sab & 15.22\\
NGC 4686				 &	XBONG	& 0.0167	&      -       & 43.24 & 23.64 &    -  & 2.84 & - & 0.45 & Sa & - \\
NGC 4992				 &	XBONG	& 0.0251	&   0.069 & 43.89 & 23.69 & -  & 2.72 & 0.13 & 0.35 & Sa & 45.99 \\
NGC 5610				 &	Elusive	& 0.0168	&   0.052 & 43.09 & 22.66 & 0.458 & 2.26 & - & 0.2 & SBab & 26.50 \\
UGC 05881				 & 	Elusive	& 0.0206   &   0.045 & 43.27 &  22.93 & 0.249 & 2.40 & 0.12 & 0.16 & Sa & 23.32 \\

 \enddata
 
 \tablecomments{XBONGs / optically elusive AGN and their classification, redshift, Eddington Ratio, X-ray luminosities, X-ray column densities, $E(B-V)$ reddening, $U-R$ color, lower limit on the SFR from \emph{GALEX} FUV fluxes, the ratio of the light in a 1.5\arcsec~radius compared to light in a 50\arcsec radius of each galaxy, the morphological type, and the strength of the 4000~\AA~break as defined by Equation~1, Section \ref{sec:4000a}. Note that the SDSS fiber width is 3\arcsec, so the final column gives the fraction of total galaxy light that is integrated to produce the SDSS spectrum. The final column is the galaxy's morphological classification in the HyperLeda database, or, for the first three objects, our own classification. A dash indicates that the quantity could not be measured due to lack of data, or is not relevant for object class.}

 \end{deluxetable}
\newpage

 \begin{deluxetable}{lcccccccc}
 \label{tab:list}
  \tabletypesize{\footnotesize}
 \tablewidth{0pt}
 \tablecaption{Infrared AGN Indicators\label{t:tab2}}
 \tablehead{
 \colhead{Object} &
 \colhead{Class} &
 \colhead{IR} &
 \colhead{NIR Coronal} &
 \colhead{Hidden} &
 \colhead{\nev} &
 \colhead{\oiv} &
 \colhead{IR} &
 \colhead{Prominent Hot} \\
 \colhead{ } &
 \colhead{ } &
 \colhead{Lines} &
 \colhead{Lines} &
 \colhead{BLR} &
 \colhead{Detection} &
 \colhead{Detection} &
 \colhead{Photometry } &
 \colhead{Dust Component}}
 \startdata 
 
2MASX J1439+1415		 &     XBONG	& F	&  F & F &  - & - &	T$^{1,2,3}$ & T \\
2MFGC 00829      			 &	XBONG	& F 	&  F & F &  - &- & F & F \\
KUG 1238+278A			 &	Elusive	& T	&  F & T & - & - &T$^{1,2,3}$ & T \\
Mrk 18					 &	Elusive	& T	&  F & F & T & T & F & T \\
NGC 4686				 &	XBONG	& F	&  F & F &  F &T & F & F \\
NGC 4992				 &	XBONG	& F	&  F & F & F & T & T$^{1,2}$ & F\\
NGC 5610				 &	Elusive	& T	&   F & F & - & - & F & T \\
UGC 05881				 & 	Elusive	& T   &   F & F & - & - & F & F\\

 \enddata
 
 \tablecomments{Results of various tests of AGN indication. An $F$ indicates the object is not identified as an AGN by the test, a $T$ indicates that the object was classified as an AGN, and a dash indicates that the test could not be performed on the object. Indices in the IR Photometry column indicate which of the MIR AGN classification schemes the object passed: (1) \citet{stern12}, (2) \citet{mateos12}, or (3) \citet{edelson12}.}

 \end{deluxetable}

\newpage


 \begin{deluxetable}{lcccccccccc}
 \label{tab:list}
  \tabletypesize{\scriptsize}
  \tabcolsep=0.10cm
 \tablewidth{0pt}
 \tablecaption{Line Fluxes and Upper Limits\label{t:tab3}}
 \tablehead{
 \colhead{Object} &
 \colhead{H$\beta$} &
 \colhead{\oiii} &
  \colhead{\oi} &
 \colhead{H$\alpha$} &
 \colhead{\nii} &
 \colhead{\sii} &
 \colhead{\feii} &
 \colhead{Pa$\beta$} &
 \colhead{H$_2$} &
 \colhead{Br$\gamma$} \\
 \colhead{ } &
 \colhead{ 4861 \AA } &
 \colhead{ 5007 \AA } &
 \colhead{ 6300 \AA } &
 \colhead{ 6563 \AA } &
 \colhead{ 6584 \AA } &
 \colhead{ 6717,31 \AA } &
 \colhead{ 1.257 $\mu$m } &
 \colhead{ 1.282 $\mu$m } &
 \colhead{ 2.121 $\mu$m } &
 \colhead{ 2.165 $\mu$m }} 

 \startdata 
 
2MASX J1439+1415 &  $<$0.45 & $<$0.52 & $<$0.11 & 5.69$\pm$0.01 & 1.41$\pm$0.09 & $<$2.80 & $<$1.08 & - & $<$0.27 & $<$1.81 \\
2MFGC 00829          &  $<$1.23 &  8.21$\pm$0.53 & $<$0.56 & $<$2.42 & $<$5.33 & $<$6.82  & $<$1.149 & - & $<$0.59 & $<$0.63 \\
KUG 1238+278A	&	4.59$\pm$0.04	& 5.80$\pm$0.05 & 2.89$\pm0.07$ &  30.77$\pm$0.06 & 20.22$\pm$0.09 & 13.12$\pm0.14$ & 2.75$\pm$0.08 & 10.40$\pm$0.15 & 1.32$\pm$0.13 & 2.21$\pm$0.19  \\
Mrk 18			&17.61$\pm$0.15	& 24.76$\pm$0.21 & 3.25$\pm0.29$ &  56.56$\pm$0.24 & 30.51$\pm$0.24 & 27.44$\pm0.51$ & 3.40$\pm$0.34 & 5.39$\pm$0.48 & 1.34$\pm$0.93 & 1.18$\pm$0.52\\
NGC 4686                  & $<$7.23 & $<$5.36 & - & - & - & - & $<$2.24 & $<$2.16 & $<$1.55 & $<$1.01\\
NGC 4992                  & $<$1.70 & $<$3.1 & $<$0.79 & 1.42$\pm$0.25 & 6.10$\pm$0.30 &$<$4.10 & $<$1.18 & $<$0.84 & $<$0.58 & $<$0.34 \\
NGC 5610		&7.26$\pm$0.07 & 7.17$\pm$0.08	& 2.34$\pm0.12$ &  50.10$\pm$0.11 & 35.71$\pm$0.10 & 20.33$\pm0.21$ & 6.90$\pm$0.11 & 11.90$\pm$0.12 & 3.27$\pm$0.19 & 4.75$\pm$0.21\\
UGC 05881		 & 15.35$\pm$0.19	& 25.43$\pm$0.18   &  2.54$\pm0.29$ & 77.35$\pm$0.33 & 47.03$\pm$0.22 & 25.70$\pm0.58$ &  5.96$\pm$0.04 & 7.78$\pm$0.39 & 2.16$\pm$0.74 & 2.07$\pm$0.65\\
	
 \enddata
 
 \tablecomments{~Line fluxes or 2-$\sigma$ upper limits for the optical lines used in the \citet{kewley06} diagnostics and infrared lines used in the \citet{rodrig04} and \citet{riffel13} diagnostics for the optically elusive AGN, with 1-$\sigma$ errors. A dash indicates that the quantity was unmeasurable due to poor spectral quality or telluric contamination. Flux units are are $10^{-15}$ erg cm$^{-2}$s$^{-1}$.}

 \end{deluxetable}

\newpage


\end{document}